\documentclass[12pt,draftcls,onecolumn]{IEEEtran}
\usepackage{cite}
\ifCLASSINFOpdf
\else
  \usepackage[dvips]{graphicx}
  \graphicspath{{../Plots/}}
  \DeclareGraphicsExtensions{.eps}
\fi
\usepackage[cmex10]{amsmath}
\usepackage{amssymb}
\usepackage{array}

\usepackage[tight,footnotesize]{subfigure}

\usepackage[font=footnotesize]{subfig}

\usepackage{pstricks-add}

\DeclareMathAlphabet{\mathbit}{OML}{cmr}{bx}{it}

\DeclareMathOperator{\E}{E}
\DeclareMathOperator{\Probability}{Pr}

\newcommand{\ve}[1]{\mbox{$\mathbit#1$}}
\newcommand{\loga}[1]{\log{\left(#1\right)}}
\newcommand{\norm}[1]{\left|#1\right|}
\newcommand{\ex}[1]{\E \left[#1\right]}
\newcommand{\Prob}[1]{\Probability\left[#1\right]}

\newtheorem{theorem}{Theorem}

\newenvironment{proof}[1][Proof]{\begin{trivlist}
\item[\hskip \labelsep {\bfseries #1}]}{\end{trivlist}}

\begin{document}
%
\title{Towards Optimal Schemes for the\\Half-Duplex Two-Way Relay Channel}
%
%
%

\author{Manuel Stein\\Technische Universit\"at M\"unchen, Germany\\
E-Mail: \{manuel.stein\}@tum.de}

\IEEEspecialpapernotice{(Technical Note, August 2011)}

\maketitle
\pagestyle{empty}
\thispagestyle{empty}
\begin{abstract}
A restricted two-way communication problem in a small fully-connected network is investigated. The network consists of three nodes, all having access to a common channel with half-duplex constraint. Two nodes want to establish a dialog while the third node can assist in the bi-directional transmission process. All nodes have agreed on a transmission protocol a priori and the problem is restricted to the dialog encoders not being allowed to establish a cooperation by the use of previous receive signals. The channel is referred to as the \emph{restricted half-duplex two-way relay channel}. Here the channel is defined and an outer bound on the achievable rates is derived by the application of the \emph{cut-set theorem}. This shows that the problem consists of six parts. We propose a transmission protocol which takes into account all possible transmit-receive configurations of the network and performs partial decoding of the messages at the relay as well as sequential decoding at the dialog nodes. By the use of random codes and suboptimal decoders, two inner bound on the achievable rates are derived. Restricting to the suggested strategies and fixed input distributions it is argued to be possible to determine optimal transmission schemes with respect to various reasonable objectives at low complexity. In comparison to two-way communication without relay, simulations for an AWGN channel model then show that it is possible to simultaneously increase the communication rates of both dialog messages and to outperform relaying strategies that ignore an available direct path.
\end{abstract}


\IEEEpeerreviewmaketitle

\section{Introduction}
\IEEEPARstart{I}{}n the last years it has been recognized that operating wireless systems in a point-to-point fashion is not optimal. Other users act as interferers in the transmission process and are therefore equivalent to noise. Cooperation has shown to be capable of dominating this competitive approach. A cooperative concept, that is already considered for upcoming wireless standards and has therefore received increasing attention lately, is known as \emph{relaying}. Source and destination connect over one or many intermediate nodes if isolated from each other or when facing bad channel conditions for direct communication. The potential to further extend the efficient use of resources when interchanging data between two nodes in a bi-directional way over a relay \cite{Rank05} has led to a variety of recent works. Although, some years ago, it has been shown that with a careful design of transmission protocols \cite{Cov79}, one-way relaying can also be used to increase the communication rate in the presence of a direct path, two-way relaying is still mostly considered in the context of a connectivity problem, e.g., \cite{Rank05} \cite{Oech08} \cite{Guend08} \cite{Schnurr07} \cite{Nam08} \cite{Nam10} \cite{Fon11} \cite{Wil10}. Two nodes cannot establish a direct communication and therefore convey messages over a third node, the relay.  A more general approach to full-duplex two-way relaying was suggested in \cite{Rank06}. Two nodes that have a direct connection want to perform a dialog. It is to decide if a third node should join the communication process in order to facilitate the exchange of messages. The motivation to do so could be to maximize the communication rates for given resources or to minimize resources for given rates. The ultimate solution to both problems is to use the channel in the most efficient way by organizing the communication process in the network.\\\\
Here we study the fully-connected problem \cite{Rank06} under a half-duplex constraint which is reasonable due to the ability of todays radio hardware in wireless systems. The current discussion on two-way relaying focuses on the topic in order to increase the service for users in bad channel conditions, e.g., a mobile user at the cell edge of a wireless system. Here we show that it is possible to significantly increase the communication rates of half-duplex constraint users by the use of two-way relaying even when good connectivity is available. The key is to allow the usage of all transmission possibilities in the system and then weight them according to the considered objective. This generic approach includes the relaying situation when no direct path is available as special case.
\subsection{Outline and Notation}
In order to motivate the considered problem we start with a toy example on a simplified network model with orthogonal and symmetric links. After summarizing related works, a definition of the communication problem is given. The \emph{cut-set theorem} is applied in order to derive an outer bound on the achievable rate pairs. Then two inner bounds with decoding at the relay are given. The more general strategy, achieving capacity on the toy example, is derived through a transparent proof technique. To make full use of the channel, message indices are mapped (or \emph{reindexed}) to multidimensional index structures which allow to distribute the messages over different transmission possibilities offered by the network. Sequential decoding at the dialog nodes disciplines the derivation of rate constraints under which error-free transmission for long codes is possible. The second inner bound follows directly as a special case of the first bound. Then we comment on the problem of finding optimal schemes with respect to various objectives under the given strategies and it's relevance for wireless communication systems. Finally, the focus lies on a scalar \emph{AWGN channel} model for which the scheme that maximizes the symmetric two-way communication rate with full decoding at the relay is determined and the performance gain is visualized.\\\\
In the following $P_{X}(\cdot)$ denotes the probability distribution of a random variable $X$ with finite and discrete alphabet $\mathcal{X}$. $P_{Y|X}(\cdot|\cdot)$ denotes the probability distribution of $Y$ given $X$.  For brevity we set aside the subscript were the associated variables become clear from the context.  A certain realization of the random variable $X$ is denoted $x$. A sequence of $n$ random variables is indicated by $X^n$. The $k$-th variable in $X^n$ is addressed by $X_{k}$. $I(X;Y)$ is the mutual information between $X$ and $Y$. The $\epsilon$-$letter$ $typical$ $set$ $T_{\epsilon}^n(P_{XY})$ with respect to $P_{XY}(\cdot)$ is defined as \cite{Kra07}
\begin{align}
T_{\epsilon}^n(P_{XY})=\left\{(x^n,y^n):\norm{ \frac{1}{n} N(a,b|x^n,y^n)-P_{XY}(a,b)}\leq\epsilon P_{XY}(a,b),\forall (a,b)\in\mathcal{X}\times\mathcal{Y}\right\}
\end{align}
where $N(a,b|x^n,y^n)$ is the number of occurrences of the pair $(a,b)$ in the sequence $(x^n,y^n)$ and $\epsilon>0$. The probability of $X^n$ emitted from an independent source $P_{X}(\cdot)$ falling into $T_{\epsilon}^n(P_{XY})$ for a given sequence $y^n\in T_{\epsilon'}^n(P_{Y})$ with $0\leq\epsilon'<\epsilon$ can be upper bounded by
\begin{align}
\Prob{X^n \in T_{\epsilon}^n(P_{XY})}\leq 2^{-n[I(X;Y)-2 \epsilon H(X)]}.
\end{align}
\subsection{Wireline Example}
\begin{figure}[!ht]
\centering
\psset{yunit=0.15cm,xunit= 0.15cm,runit= 0.15cm}
\subfigure{
\begin{pspicture}[showpoints=false](2,2)(18,14)
	
\cnode(4,4){1.5}{N1} 
\cnode(10,12){1.5}{N2}
\cnode(16,4){1.5}{N3}
	
\rput(4,4){\small{$1$}}
\rput(10,12){\small{$2$}}
\rput(16,4){\small{$3$}}
	
\rput(5.5,10){\small{$b_1$}}
\rput(10,1.5){\small{$b_2$}}

\ncarc[arcangle=10,nodesep=2pt]{->}{N1}{N2}
\ncarc[arcangle=-10,nodesep=2pt]{->}{N1}{N3}
\ncarc[arcangle=10,linestyle=dotted,nodesep=2pt]{-}{N2}{N3}
	
\end{pspicture}}
%
%
\subfigure{
\begin{pspicture}[showpoints=false](2,2)(18,14)
	
\cnode(4,4){1.5}{N1} 
\cnode(10,12){1.5}{N2}
\cnode(16,4){1.5}{N3}
	
\rput(4,4){\small{$1$}}
\rput(10,12){\small{$2$}}
\rput(16,4){\small{$3$}}

\rput(14.5,10){\small{$b_3$}}
\rput(10,1.5){\small{$b_4$}}
		
\ncarc[arcangle=-10,nodesep=2pt]{->}{N3}{N2}
\ncarc[arcangle=10,nodesep=2pt]{->}{N3}{N1}
\ncarc[arcangle=-10,linestyle=dotted,nodesep=2pt]{-}{N2}{N1}
	
\end{pspicture}}
%
%
\subfigure{
\begin{pspicture}[showpoints=false](2,2)(18,14)
	
\cnode(4,4){1.5}{N1} 
\cnode(10,12){1.5}{N2}
\cnode(16,4){1.5}{N3}
	
\rput(4,4){\small{$1$}}
\rput(10,12){\small{$2$}}
\rput(16,4){\small{$3$}}

\rput(5.5,10){\small{$b_3$}}
\rput(14.5,10){\small{$b_1$}}

\ncarc[arcangle=-10,nodesep=2pt]{->}{N2}{N1} 
\ncarc[arcangle=10,nodesep=2pt]{->}{N2}{N3} 
\ncarc[arcangle=-10,linestyle=dotted,nodesep=2pt]{-}{N1}{N3}

\end{pspicture}}
\caption{Alternative three-step scheme (1.33 bps / 1.5 lpb / 0.75 npb)}
\label{fig:2WRC3STEPS_ALT}
\end{figure}
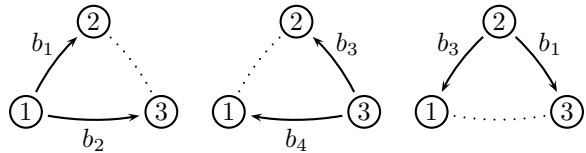

\begin{figure}[!ht]
\centering
\psset{yunit=0.125cm,xunit= 0.125cm,runit= 0.125cm}
\subfigure{
\begin{pspicture}[showpoints=false](3,2)(17,14)
	
\cnode(4,4){1.5}{N1} 
\cnode(10,12){1.5}{N2}
\cnode(16,4){1.5}{N3}
	
\rput(4,4){\small{$1$}}
\rput(10,12){\small{$2$}}
\rput(16,4){\small{$3$}}
	
\rput(5.5,10){\small{$b_1$}}
\rput(10,1.5){\small{$b_2$}}

\ncarc[arcangle=10,nodesep=2pt]{->}{N1}{N2}
\ncarc[arcangle=-10,nodesep=2pt]{->}{N1}{N3}
\ncarc[arcangle=10,linestyle=dotted,nodesep=2pt]{-}{N2}{N3}
	
\end{pspicture}}
%
%
\subfigure{
\begin{pspicture}[showpoints=false](3,2)(17,14)
	
\cnode(4,4){1.5}{N1} 
\cnode(10,12){1.5}{N2}
\cnode(16,4){1.5}{N3}
	
\rput(4,4){\small{$1$}}
\rput(10,12){\small{$2$}}
\rput(16,4){\small{$3$}}

\rput(14.5,10){\small{$b_1$}}
\rput(10,1.5){\small{$b_3$}}
		
\ncarc[arcangle=10,nodesep=2pt]{->}{N2}{N3} 
\ncarc[arcangle=-10,nodesep=2pt]{->}{N1}{N3}
\ncarc[arcangle=10,linestyle=dotted,nodesep=2pt]{-}{N1}{N2}

\end{pspicture}}
%
%
\subfigure{
\begin{pspicture}[showpoints=false](3,2)(17,14)
	
\cnode(4,4){1.5}{N1} 
\cnode(10,12){1.5}{N2}
\cnode(16,4){1.5}{N3}
	
\rput(4,4){\small{$1$}}
\rput(10,12){\small{$2$}}
\rput(16,4){\small{$3$}}

\rput(14.5,10){\small{$b_4$}}
\rput(10,1.5){\small{$b_5$}}
		
\ncarc[arcangle=-10,nodesep=2pt]{->}{N3}{N2}
\ncarc[arcangle=10,nodesep=2pt]{->}{N3}{N1}
\ncarc[arcangle=-10,linestyle=dotted,nodesep=2pt]{-}{N2}{N1}
	
\end{pspicture}}
%
%
\subfigure{
\begin{pspicture}[showpoints=false](3,2)(17,14)
	
\cnode(4,4){1.5}{N1} 
\cnode(10,12){1.5}{N2}
\cnode(16,4){1.5}{N3}
	
\rput(4,4){\small{$1$}}
\rput(10,12){\small{$2$}}
\rput(16,4){\small{$3$}}

\rput(5.5,10){\small{$b_4$}}
\rput(10,1.5){\small{$b_6$}}

\ncarc[arcangle=-10,nodesep=2pt]{->}{N2}{N1} 
\ncarc[arcangle=10,nodesep=2pt]{->}{N3}{N1} 
\ncarc[arcangle=10,linestyle=dotted,nodesep=2pt]{-}{N2}{N3}

\end{pspicture}}
\caption{Four-step scheme (1.5 bps / 1.33 lpb / 1.0 npb)}
\label{fig:2WRC4STEPS}
\end{figure}
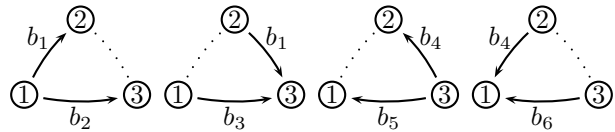

Consider a fully-connected wireline network with three nodes. Each node faces a half-duplex constraint, i.e., it can not receive and transmit simultaneously. For simplicity all links support transmission of one reliable bit per channel use. Nodes 1 and 3 want to exchange messages (bits) while node 2 can assist the communication process as a relay.
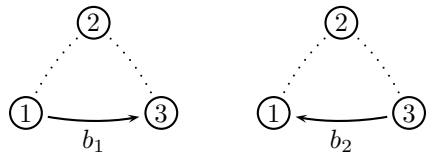
\begin{figure}[!ht]
\centering
\psset{yunit=0.15cm,xunit= 0.15cm,runit= 0.15cm}
\subfigure{
\begin{pspicture}[showpoints=false](0,2)(20,14)
	
\cnode(4,4){1.5}{N1} 
\cnode(10,12){1.5}{N2}
\cnode(16,4){1.5}{N3}
	
\rput(4,4){\small{$1$}}
\rput(10,12){\small{$2$}}
\rput(16,4){\small{$3$}}
	
\rput(10,1.5){\small{$b_1$}}

\ncarc[arcangle=10,linestyle=dotted,nodesep=2pt]{-}{N1}{N2}
\ncarc[arcangle=-10,nodesep=2pt]{->}{N1}{N3}
\ncarc[arcangle=10,linestyle=dotted,nodesep=2pt]{-}{N2}{N3}

\end{pspicture}}
%
%
\subfigure{
\begin{pspicture}[showpoints=false](0,2)(20,14)
	
\cnode(4,4){1.5}{N1} 
\cnode(10,12){1.5}{N2}
\cnode(16,4){1.5}{N3}
	
\rput(4,4){\small{$1$}}
\rput(10,12){\small{$2$}}
\rput(16,4){\small{$3$}}

\rput(10,1.5){\small{$b_2$}}
		
\ncarc[arcangle=10,linestyle=dotted,nodesep=2pt]{-}{N2}{N3} 
\ncarc[arcangle=10,nodesep=2pt]{->}{N3}{N1}
\ncarc[arcangle=10,linestyle=dotted,nodesep=2pt]{-}{N1}{N2}

\end{pspicture}}
\caption{Two-way channel (1.0 bps / 1.0 lpb / 1.0 npb)}
\label{fig:2WRCTWC}
\end{figure}
The simplest way to communicate is to let the dialog nodes send sequentially to each other while node 2 stays turned off (see Fig. \ref{fig:2WRCTWC}). This communication strategy, referred to as two-way channel (TWC), allows to transmit two bits within two steps. The two-way communication rate is $R=1.0$ bits per step (bps). Asking for the cost $C_{T}$ of the communication protocol under simplified assumptions we can conclude that two links have been used, $C_{T}=1.0$ links per bit (lpb) or, as an alternative simple cost model, two nodes have been activated, $C_{T}=1.0$ nodes per bit (npb).
%
%
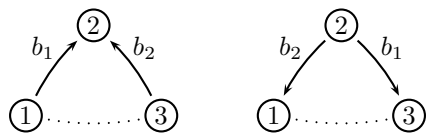
\begin{figure}[!ht]
\centering
\psset{yunit=0.15cm,xunit= 0.15cm,runit= 0.15cm}
%
%
\subfigure{
\begin{pspicture}[showpoints=false](0,2)(20,14)
	
\cnode(4,4){1.5}{N1} 
\cnode(10,12){1.5}{N2}
\cnode(16,4){1.5}{N3}
	
\rput(4,4){\small{$1$}}
\rput(10,12){\small{$2$}}
\rput(16,4){\small{$3$}}

\rput(5.5,10){\small{$b_1$}}
\rput(14.5,10){\small{$b_2$}}

\ncarc[arcangle=10,nodesep=2pt]{->}{N1}{N2} 
\ncarc[arcangle=-10,nodesep=2pt]{->}{N3}{N2} 
\ncarc[arcangle=-10,linestyle=dotted,nodesep=2pt]{-}{N1}{N3}

\end{pspicture}}
%
%
\subfigure{
\begin{pspicture}[showpoints=false](0,2)(20,14)
	
\cnode(4,4){1.5}{N1} 
\cnode(10,12){1.5}{N2}
\cnode(16,4){1.5}{N3}
	
\rput(4,4){\small{$1$}}
\rput(10,12){\small{$2$}}
\rput(16,4){\small{$3$}}

\rput(5.5,10){\small{$b_2$}}
\rput(14.5,10){\small{$b_1$}}

\ncarc[arcangle=-10,nodesep=2pt]{->}{N2}{N1} 
\ncarc[arcangle=10,nodesep=2pt]{->}{N2}{N3} 
\ncarc[arcangle=-10,linestyle=dotted,nodesep=2pt]{-}{N1}{N3}

\end{pspicture}}
\caption{Two-step scheme (1.0 bps / 2.0 lpb / 1.5 npb)}
\label{fig:2WRC2STEPS}
\end{figure}
A popular two-way relaying scheme \cite{Rank05} is provided by the two steps of Fig. \ref{fig:2WRC2STEPS}. For this scheme the rate stays at $R=1.0$ bps while the cost increases to $C_{T}=2.0$ lpb or $C_{T}=1.5$ npb. At a higher cost the same rate is attained like with the TWC (Fig. \ref{fig:2WRCTWC}). Note that this scheme ignores the possibility to use the direct path between node 1 and node 3. 
%
%
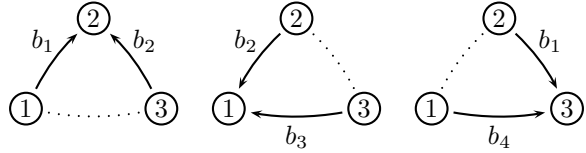
\begin{figure}[!ht]
\centering
\psset{yunit=0.15cm,xunit= 0.15cm,runit= 0.15cm}
\subfigure{
\begin{pspicture}[showpoints=false](2,2)(18,14)
	
\cnode(4,4){1.5}{N1} 
\cnode(10,12){1.5}{N2}
\cnode(16,4){1.5}{N3}
	
\rput(4,4){\small{$1$}}
\rput(10,12){\small{$2$}}
\rput(16,4){\small{$3$}}
	
\rput(5.5,10){\small{$b_1$}}
\rput(14.5,10){\small{$b_2$}}

\ncarc[arcangle=10,nodesep=2pt]{->}{N1}{N2}
\ncarc[arcangle=-10,nodesep=2pt]{->}{N3}{N2}
\ncarc[arcangle=-10,linestyle=dotted,nodesep=2pt]{-}{N1}{N3}
	
\end{pspicture}}
%
%
\subfigure{
\begin{pspicture}[showpoints=false](2,2)(18,14)
	
\cnode(4,4){1.5}{N1} 
\cnode(10,12){1.5}{N2}
\cnode(16,4){1.5}{N3}
	
\rput(4,4){\small{$1$}}
\rput(10,12){\small{$2$}}
\rput(16,4){\small{$3$}}

\rput(5.5,10){\small{$b_2$}}
\rput(10,1.5){\small{$b_3$}}

\ncarc[arcangle=-10,nodesep=2pt]{->}{N2}{N1} 
\ncarc[arcangle=10,nodesep=2pt]{->}{N3}{N1} 
\ncarc[arcangle=10,linestyle=dotted,nodesep=2pt]{-}{N2}{N3}

\end{pspicture}}
%
%
\subfigure{
\begin{pspicture}[showpoints=false](2,2)(18,14)
	
\cnode(4,4){1.5}{N1} 
\cnode(10,12){1.5}{N2}
\cnode(16,4){1.5}{N3}
	
\rput(4,4){\small{$1$}}
\rput(10,12){\small{$2$}}
\rput(16,4){\small{$3$}}

\rput(14.5,10){\small{$b_1$}}
\rput(10,1.5){\small{$b_4$}}
		
\ncarc[arcangle=10,nodesep=2pt]{->}{N2}{N3}
\ncarc[arcangle=-10,nodesep=2pt]{->}{N1}{N3}
\ncarc[arcangle=10,linestyle=dotted,nodesep=2pt]{-}{N1}{N2}
	
\end{pspicture}}
\caption{Three-step scheme (1.33 bps / 1.5 lpb / 1.5 npb)}
\label{fig:2WRC3STEPS}
\end{figure}
An improvement in transmission rate is provided by the three-step scheme sketched in Fig. \ref{fig:2WRC3STEPS}. The rate attains $R=1.33$ bps while the cost is $C_{T}=1.5$ lpb or $C_{T}=1.5$ npb. In order to activate less nodes in the communication process, one could think of a different three-step scheme \cite{Pop07} (see Fig. \ref{fig:2WRC3STEPS_ALT}). For this protocol the rate stays at $R=1.33$ bps while the cost is $C_{T}=1.5$ lpb or $C_{T}=0.75$ npb. Note that interestingly with this scheme the rate is higher than for the simple two-way communication without relay at a lower node-based cost. 
%
%
Finally, consider a four-step scheme like depicted in Fig. \ref{fig:2WRC4STEPS}. The rate has reached $R=1.5$ bps at a cost of $C_{T}=1.33$ lpb or $C_{T}=1.0$ npb.
%
%
\subsection{Potentials and Critics}
The given example indicates that two-way relaying schemes offer possibilities to enhance the performance of bi-directional communication on half-duplex systems even in the presence of a direct connection. This is in particular interesting for wireless communication scenarios. For example an idle user could join the communication process between a base-station and a mobile in order to allow higher transmission rates. Obviously, such examples with orthogonal links neglect the properties of wireless channels, i.e., links supporting asymmetric rates, statistical dependence of links resulting in broadcasting and superposition. However, note that there are wireless channels that allow to enforce situations similar to wireline through \emph{pre-coding techniques}, e.g., MIMO wireless channels \cite{Lo05}. The example also suggests that treating the fully-connected problem like a \emph{separated two-way relay channel} \cite{Guend08} and ignoring the direct path is not optimal. This raises the question: Which half-duplex scheme is the optimal one for the fully-connected problem sketched here? All the following is a first step towards an answer to this particular question. 
\section{Related Work}
The relay channel was introduced in \cite{Meul71}. The seminal work \cite{Cov79} derives an upper bound on the capacity of the relay channel by developing cut-set arguments. Moreover, different relaying strategies are presented, among them the \emph{decode-and-forward} strategy which is shown to be capacity achieving for the \emph{degraded relay channel} \cite[Theorem 1]{Cov79}. In \cite[Section 9]{Kra07} relaying strategies are comprehensively reviewed and a \emph{partial-decode-and-forward} strategy is proposed. This forms a special case of \cite[Theorem 7]{Cov79} and achieves capacity on the \emph{semideterministic} relay channel \cite{Gam82} and a relay channel with \emph{orthogonal components} \cite{Gam05}. The half-duplex relay channel is explored in \cite{Ho05}. Two-way communication channels were introduced in \cite{Shan61}. The work \cite{Rank05} establishes the idea of exchanging messages in a bi-directional way over a relay. In \cite{Oech08} the broadcast phase of the \emph{two-phase two-way relay channel} (see Fig. \ref{fig:2WRC2STEPS}) is considered and the capacity achieving coding scheme is derived if the relay has available both dialog messages. Using a \emph{compress-and-forward} strategy for this relay channel is analyzed in \cite{Schnurr07} while \cite{Nam08} \cite{Nam10} and \cite{Wil10} show how to outperform this by using \emph{structured codes}. \cite{Pop07} proposes a three-phase scheme (see Fig. \ref{fig:2WRC3STEPS_ALT}) and obtains the achievable rates with network coding. This scheme is also examined in \cite{Schnurr08} and additionally a four-phase scheme is put forward. \cite{Rank06} studies the full-duplex two-way relay channel while \cite{Guend08} focuses on a separated full-duplex model. A separated two-way relay channel with feedback is subject of the work \cite{Fon11}. \cite{Ave08} comes up with a deterministic approach to approximate the capacity of the Gaussian two-way relay channel.
\section{Communication Problem}
\begin{figure}[!ht]
\centering
\psset{yunit=0.21cm,xunit= 0.21cm,runit= 0.21cm}
\begin{pspicture}[showpoints=false](0,-1)(40,13.5)
	
\psframe[linewidth=1pt,framearc=.1,fillstyle=solid,fillcolor=lightgray](0,0)(8,5)
\psframe[linewidth=1pt,framearc=.1,fillstyle=solid,fillcolor=lightgray](16,8.5)(24,13.5)
\psframe[linewidth=1pt,framearc=.1,fillstyle=solid,fillcolor=lightgray](32,0)(40,5)
			
\uput{0 ex}[90](4,2.0){\small{Node 1}}
\uput{0 ex}[90](20,10.5){\small{Node 2}}
\uput{0 ex}[90](36,2.0){\small{Node 3}}
			
\psframe[linewidth=1pt,framearc=.1,fillstyle=solid,fillcolor=lightgray](11.5,0)(28.5,5)
\uput{0 ex}[90](20.1,1.8){\small{$P(y_1 y_2 y_3|x_1 x_2 x_3 s)$}}
			
\psline[linewidth=1pt]{->}(8,4)(11.5,4)
\psline[linewidth=1pt]{<-}(8,1)(11.5,1)
\psline[linewidth=1pt]{->}(2,7)(2,5)
\psline[linewidth=1pt]{->}(6,5)(6,7)
			
\uput{0.5 ex}[90](2,7){\small{$W_{1\text{-}3}$}}
\uput{0.5 ex}[90](6,7){\small{$\hat{W}_{3\text{-}1}$}}
\uput{0.5 ex}[90](9.75,4){\small{$X_{1}$}}
\uput{0.5 ex}[270](9.75,1){\small{$Y_{1}$}}
			
\psline[linewidth=1pt]{->}(18,8.5)(18,5)
\psline[linewidth=1pt]{<-}(22,8.5)(22,5)
	
\uput{0.5 ex}[180](18,6.75){\small{$X_{2}$}}
\uput{0.5 ex}[0](22,6.75){\small{$Y_{2}$}}
			
\psline[linewidth=1pt]{<-}(28.5,4)(32,4)
\psline[linewidth=1pt]{->}(28.5,1)(32,1)
\psline[linewidth=1pt]{->}(34,5)(34,7)
\psline[linewidth=1pt]{<-}(38,5)(38,7)
			
\uput{0.5 ex}[90](38,7){\small{$W_{3\text{-}1}$}}
\uput{0.5 ex}[90](34,7){\small{$\hat{W}_{1\text{-}3}$}}
\uput{0.5 ex}[90](30.25,4){\small{$X_{3}$}}
\uput{0.5 ex}[270](30.25,1){\small{$Y_{3}$}}
\end{pspicture}
\caption{Half-Duplex Two-Way Relay Channel}
\label{fig:HDTWRC}
\end{figure}
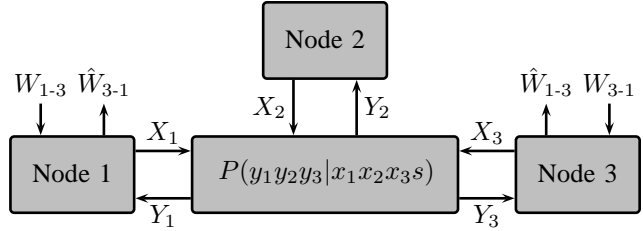

The studied network consists of three nodes labeled by $i=1, 2, 3$. The message $W_{1\text{-}3}$ is to be communicated from node 1 to node 3 and the message $W_{3\text{-}1}$ from node 3 to node 1. Node 2 has no own message but can help in the bi-directional communication process. The messages $W_{1\text{-}3}$ and $W_{3\text{-}1}$ are considered to be independent and drawn from uniform distributions. Each node $i$ is equipped with an input $X_i$ to and an output $Y_i$ from a common channel (see Fig. \ref{fig:HDTWRC}).
%
%
The channel is discrete and memoryless. In contrast to the full-duplex model \cite{Rank06}, a network state variable $S$ determines the receive-transmit configurations of the network nodes. Therefore, the channel can be defined
\begin{align}
&\left(\mathcal{X}_{1}\times\mathcal{X}_{2}\times\mathcal{X}_{3},P(y_1 y_2 y_3|x_1 x_2 x_3 s),\mathcal{Y}_{1}\times\mathcal{Y}_{2}\times\mathcal{Y}_{3},\mathcal{S}\right)
\end{align}
where $\mathcal{X}_{i},\mathcal{Y}_{i}$ are finite input and output alphabets. The network state variable $S:=(S_1,S_2,S_3)$ with $S_i\in\{0,1\}$, imposes the following restrictions on the output and input variables
\begin{align}
X_i&=\emptyset&\text{if}\hspace{1cm}&s_i=0\notag\\
Y_i&= \emptyset&\text{if}\hspace{1cm}&s_i=1,\hspace{1cm} i=1,2,3
\end{align}
with "$\emptyset$" being understood as a deactivation symbol. As the network state is determined by three binary variables the alphabet $\mathcal{S}$ can contain at the utmost eight symbols. In the following, it is assumed that all nodes have agreed on the network states for all $n$ channel uses of the communication protocol a priori. Note that under this assumption the transmit-receive configurations can not be used to transmit additional information like discussed in \cite{Kra04}. In order to focus on coding structure we assume fixed input distributions for each individual network state and exclude the use of \emph{time-sharing} techniques on the input distributions. Such techniques can be incorporated into the presented strategies by introducing a time-sharing variable $Q$. Further, all nodes have available the codebooks used in the network and the conditional distribution characterizing the channel. The communication process is neither limited by delay nor by complexity.
\subsection{Restricted Codes}
Consider $n$ channel uses, a certain scheme $s^n$ and a choice of $1\leq L\leq8$ used network states. The variable $s_k$ takes values in $\mathcal{S}:\{1,\ldots,L\}$ and determines the individual network state for the $k$-th of $n$ channel uses. A code of length $n$ and rates $(R_{1\text{-}3},R_{3\text{-}1})$ consists of two message sets
\begin{align}
\mathcal{W}_{1\text{-}3} &:\{1,\ldots,2^{nR_{1\text{-}3}}\}\notag\\
\mathcal{W}_{3\text{-}1} &:\{1,\ldots,2^{nR_{3\text{-}1}}\},
\end{align}
two encoding functions
\begin{align}
f_1&: \mathcal{W}_{1\text{-}3}\times\mathcal{S}^{n}\to\mathcal{X}_{1}^{n}\notag\\
f_3&: \mathcal{W}_{3\text{-}1}\times\mathcal{S}^{n}\to\mathcal{X}_{3}^{n},
\end{align}
a set of $n$ relaying functions
\begin{align}
f_{2,k}&: \mathcal{S}\times\mathcal{Y}^{k-1}\rightarrow\mathcal{X}_{2, k},\quad k=1, \ldots, n
\end{align}
and two decoding functions
\begin{align}
g_1&: \mathcal{Y}_{1}^{n}\times\mathcal{S}^{n}\times\mathcal{W}_{1\text{-}3}\to\mathcal{W}_{3\text{-}1}\notag\\
g_3&: \mathcal{Y}_{3}^{n}\times\mathcal{S}^{n}\times\mathcal{W}_{3\text{-}1}\to\mathcal{W}_{1\text{-}3}.
\end{align}
The code is restricted as the encoding functions are independent of previous receive signals. In the following $n_l$ denotes the number of occurrences of the network state $l$ in $n$ channel uses and $\tau_l$ is defined as the ratio $\tau_l=n_l/n$. The achievable rates are defined as all rate pairs $(R_{1\text{-}3},R_{3\text{-}1})$ for which a code can be constructed that allows to approach $(R_{1\text{-}3},R_{3\text{-}1})$ arbitrarily close while the probability of an error can be made arbitrarily small. 
\section{Outer Rate Bound} 
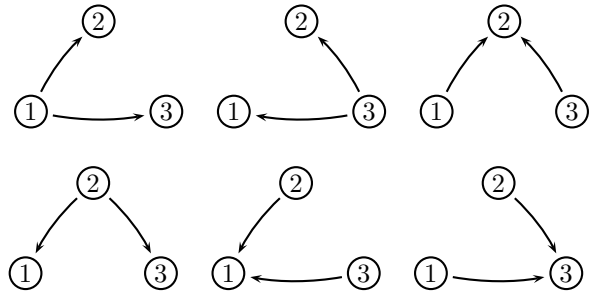
\begin{figure}[!ht]
\centering
\psset{yunit=0.15cm,xunit= 0.15cm,runit= 0.15cm}
%
%
\subfigure{
\begin{pspicture}[showpoints=false](2,2)(18,14)
	
\cnode(4,4){1.5}{N1} 
\cnode(10,12){1.5}{N2}
\cnode(16,4){1.5}{N3}
	
\rput(4,4){\small{$1$}}
\rput(10,12){\small{$2$}}
\rput(16,4){\small{$3$}}
	
\ncarc[arcangle=10,nodesep=2pt]{->}{N1}{N2}
\ncarc[arcangle=-10,nodesep=2pt]{->}{N1}{N3} 
	
\end{pspicture}}
%
%
\subfigure{
\begin{pspicture}[showpoints=false](2,2)(18,14)
	
\cnode(4,4){1.5}{N1} 
\cnode(10,12){1.5}{N2}
\cnode(16,4){1.5}{N3}
	
\rput(4,4){\small{$1$}}
\rput(10,12){\small{$2$}}
\rput(16,4){\small{$3$}}
	
\ncarc[arcangle=-10,nodesep=2pt]{->}{N3}{N2}
\ncarc[arcangle=10,nodesep=2pt]{->}{N3}{N1}

\end{pspicture}}
%
%
\subfigure{
\begin{pspicture}[showpoints=false](2,2)(18,14)
	
\cnode(4,4){1.5}{N1} 
\cnode(10,12){1.5}{N2}
\cnode(16,4){1.5}{N3}
	
\rput(4,4){\small{$1$}}
\rput(10,12){\small{$2$}}
\rput(16,4){\small{$3$}}
			
\ncarc[arcangle=10,nodesep=2pt]{->}{N1}{N2}
\ncarc[arcangle=-10,nodesep=2pt]{->}{N3}{N2}
	
\end{pspicture}}

%
%
\subfigure{
\begin{pspicture}[showpoints=false](2,4)(18,14)
	
\cnode(4,4){1.5}{N1} 
\cnode(10,12){1.5}{N2}
\cnode(16,4){1.5}{N3}
	
\rput(4,4){\small{$1$}}
\rput(10,12){\small{$2$}}
\rput(16,4){\small{$3$}}
		
\ncarc[arcangle=-10,nodesep=2pt]{->}{N2}{N1} 
\ncarc[arcangle=10,nodesep=2pt]{->}{N2}{N3} 

\end{pspicture}}
%
%
\subfigure{
\begin{pspicture}[showpoints=false](2,4)(18,14)
	
\cnode(4,4){1.5}{N1} 
\cnode(10,12){1.5}{N2}
\cnode(16,4){1.5}{N3}
	
\rput(4,4){\small{$1$}}
\rput(10,12){\small{$2$}}
\rput(16,4){\small{$3$}}
		
\ncarc[arcangle=-10,nodesep=2pt]{->}{N2}{N1} 
\ncarc[arcangle=10,nodesep=2pt]{->}{N3}{N1}
	
\end{pspicture}}
%
\subfigure{
\begin{pspicture}[showpoints=false](2,4)(18,14)
	
\cnode(4,4){1.5}{N1} 
\cnode(10,12){1.5}{N2}
\cnode(16,4){1.5}{N3}
	
\rput(4,4){\small{$1$}}
\rput(10,12){\small{$2$}}
\rput(16,4){\small{$3$}}
		
\ncarc[arcangle=10,nodesep=2pt]{->}{N2}{N3} 
\ncarc[arcangle=-10,nodesep=2pt]{->}{N1}{N3} 
	
\end{pspicture}}
\vspace{0.25cm}
\caption{Relevant network states}
\label{fig:ELEMENATRIES}
\end{figure}

In order to attain a performance bound on the problem we apply the cut-set theorem.
\hfill\\
\begin{theorem}\label{prop:UB_6P}
All rate pairs of the restricted half-duplex two-way relay channel that are achievable for some joint probability distributions of the form
\begin{align*}
P(x_{1}^{(1)} y_{2}^{(1)} y_{3}^{(1)})&=P(x_{1}^{(1)})P(y_{2}^{(1)} y_{3}^{(1)}|x_{1}^{(1)}) \notag\\
P(x_{3}^{(2)} y_{1}^{(2)} y_{2}^{(2)})&=P(x_{3}^{(2)})P(y_{1}^{(2)} y_{2}^{(2)}|x_{3}^{(2)}) \notag\\
P(x_{1}^{(3)} x_{3}^{(3)} y_{2}^{(3)})&=P(x_{1}^{(3)})P(x_{3}^{(3)})P(y_{2}^{(3)}|x_{1}^{(3)} x_{3}^{(3)}) \notag\\
P(x_{2}^{(4)} y_{1}^{(4)} y_{3}^{(4)})&=P(x_{2}^{(4)})P(y_{1}^{(4)} y_{3}^{(4)}|x_{2}^{(4)}) \notag\\
P(x_{2}^{(5)} x_{3}^{(5)} y_{1}^{(5)})&=P(x_{2}^{(5)} x_{3}^{(5)})P(y_{1}^{(5)}|x_{2}^{(5)} x_{3}^{(5)})\notag\\
P(x_{1}^{(6)} x_{2}^{(6)} y_{3}^{(6)})&=P(x_{1}^{(6)} x_{2}^{(6)})P(y_{3}^{(6)}|x_{1}^{(6)} x_{2}^{(6)})
\end{align*}
must satisfy
\begin{align*}
R_{1\text{-}3}&\leq\tau_{1}I(X_{1}^{(1)};Y_{2}^{(1)}Y_{3}^{(1)}) + \tau_{3}I(X_{1}^{(3)};Y_{2}^{(3)}|X_{3}^{(3)})+\tau_{6}I(X_{1}^{(6)};Y_{3}^{(6)}|X_{2}^{(6)})\\
R_{1\text{-}3}&\leq\tau_1 I(X_{1}^{(1)};Y_{3}^{(1)}) + \tau_4 I(X_{2}^{(4)};Y_{3}^{(4)}) + \tau_6 I(X_{1}^{(6)}X_{2}^{(6)};Y_{3}^{(6)})\\
R_{3\text{-}1}&\leq\tau_{2}I(X_{3}^{(2)};Y_{1}^{(2)}Y_{2}^{(2)}) + \tau_{3}I(X_{3}^{(3)};Y_{2}^{(3)}|X_{1}^{(3)})+\tau_{5}I(X_{3}^{(5)};Y_{1}^{(5)}|X_{2}^{(5)})\notag\\
R_{3\text{-}1}&\leq\tau_2 I(X_{3}^{(2)};Y_{1}^{(2)}) + \tau_4 I(X_{2}^{(4)};Y_{1}^{(4)}) + \tau_5 I(X_{2}^{(5)}X_{3}^{(5)};Y_{1}^{(5)})
\end{align*}
where $0\leq\tau_{l}$ and $\sum_{l=1}^{6}\tau_{l}\leq1$.
\end{theorem}
\begin{proof}
Consider a full-duplex three-node network and the two possible cut-set partitions
\begin{align}
\Omega_1:& \{\text{node 1}\} && \Omega_1^{c}:\{\text{node 2}, \text{node 3}\}\notag\\
\Omega_2:& \{\text{node 1}, \text{node 2}\} && \Omega_1^{c}:\{\text{node 3}\}
\end{align}
separating node 1 and node 3. With the cut-set theorem \cite[Theorem 15.10.1]{Cov06} it holds that the achievable rates for any joint input distribution $P(x_1x_2x_3)$ are outer bounded by
\begin{align}
\Omega_1: \quad R_{1\text{-}3}&\leq I(X_{1};Y_{2}Y_{3}|X_{2}X_{3})\notag\\
 R_{3\text{-}1}&\leq I(X_{2}X_{3};Y_{1}|X_{1})\notag\\
\Omega_2: \quad R_{1\text{-}3}&\leq I(X_{1}X_{2};Y_{3}|X_{3})\notag\\
R_{3\text{-}1}&\leq I(X_{3};Y_{1}Y_{2}|X_{1}X_{2}).
\end{align}
As here the encoders are not allowed to cooperate, we restrict the distribution $P(x_1x_2x_3)$ to factorize $P(x_1x_2x_3)=P(x_2)P(x_1|x_2)P(x_3|x_2)$. Introducing a network state variable $S$ known by all nodes, taking values in $\mathcal{S}:\{1,\ldots,L\}$ and being distributed according to $P_{S}(l)=\frac{n_l}{n}=\tau_l$ the rate constraints can be written as
\begin{align}\label{eq:SUMSTATERATES}
R_{1\text{-}3}&\leq \sum_{l=1}^{L}P_{S}(l)I(X_{1}^{(l)};Y_{2}^{(l)}Y_{3}^{(l)}|X_{2}^{(l)}X_{3}^{(l)}S=l)\notag\\
R_{1\text{-}3}&\leq \sum_{l=1}^{L}P_{S}(l)I(X_{1}^{(l)}X_{2}^{(l)};Y_{3}^{(l)}|X_{3}^{(l)}S=l)\notag\\
R_{3\text{-}1}&\leq \sum_{l=1}^{L}P_{S}(l)I(X_{3}^{(l)};Y_{1}^{(l)}Y_{2}^{(l)}|X_{1}^{(l)}X_{2}^{(l)}S=l)\notag\\
R_{3\text{-}1}&\leq \sum_{l=1}^{L}P_{S}(l)I(X_{2}^{(l)}X_{3}^{(l)};Y_{1}^{(l)}|X_{1}^{(l)}S=l).
\end{align}
Agreeing to use $L=8$ network states
\begin{align}\label{eq:def_6p_states}
l=1:s&=(1, 0, 0)&l=5:s&=(0, 1, 1)\notag\\
l=2:s&=(0, 0, 1)&l=6:s&=(1, 1, 0)\notag\\
l=3:s&=(1, 0, 1)&l=7:s&=(1, 1, 1)\notag\\
l=4:s&=(0, 1, 0)&l=8:s&=(0, 0, 0)
\end{align}
establishes the theorem after a straight-forward expansion of Eq. (\ref{eq:SUMSTATERATES}).
\end{proof}
By the proof it becomes clear that the problem contains six separate parts (see Fig. \ref{fig:ELEMENATRIES}) without a specific order. Three parts where one node sends to two other nodes and three parts where two nodes send to one of the nodes.
%
%
The network configurations where all nodes transmit or receive do not offer a positive information flow in the network. Note that for any scheme which uses less than the six relevant network states (see Fig. \ref{fig:ELEMENATRIES}) an individual performance outer bound can be derived without further proof by setting the probability of the unused network states in Theorem \ref{prop:UB_6P} to zero, i.e., $P_S(l)=\tau_l=0$. 
\section{Inner Rate Bound}
In order to attain an inner bound we suggest a scheme that takes into account all the transmit-receive configurations that have been revealed to be relevant for the problem by the cut-set bound. The proposed six-phase scheme will be referred to as 6P. The phases $1$ to $6$ are defined like in Eq. (\ref{eq:def_6p_states}).
\hfill\\
\begin{theorem}\label{prop:AR_6P_PDF}
All rate pairs of the half-duplex two-way relay channel that satisfy
\begin{align*}
R_{1\text{-}3} &\leq \tau_{1}\Big(I(U_{1}^{(1)};Y_{2}^{(1)})+I(X_{1}^{(1)};Y_{3}^{(1)}|U_{1}^{(1)})\Big) + \tau_{3}I(X_{1}^{(3)};Y_{2}^{(3)}|X_{3}^{(3)})+ \tau_{6}I(X_{1}^{(6)};Y_{3}^{(6)}|X_{2}^{(6)})\\
R_{1\text{-}3} &\leq \tau_{1} I(X_{1}^{(1)};Y_{3}^{(1)}) + \tau_{4} I(X_{2}^{(4)};Y_{3}^{(4)})+ \tau_{6} I(X_{1}^{(6)}X_{2}^{(6)};Y_{3}^{(6)})\\
R_{3\text{-}1} &\leq \tau_{2}\Big(I(U_{3}^{(2)};Y_{2}^{(2)})+I(X_{3}^{(2)};Y_{1}^{(2)}|U_{3}^{(2)})\Big) + \tau_{3}I(X_{3}^{(3)};Y_{2}^{(3)}|X_{1}^{(3)})+ \tau_{5}I(X_{3}^{(5)};Y_{1}^{(5)}|X_{2}^{(5)})\\
R_{3\text{-}1} &\leq \tau_{2} I(X_{3}^{(2)};Y_{1}^{(2)}) + \tau_{4} I(X_{2}^{(4)};Y_{1}^{(4)})+\tau_{5} I(X_{2}^{(5)}X_{3}^{(5)};Y_{1}^{(5)})
\end{align*}
subject to
\begin{align*}
R_{1\text{-}3}+R_{3\text{-}1} &\leq \tau_{1}\Big(I(U_{1}^{(1)};Y_{2}^{(1)})+I(X_{1}^{(1)};Y_{3}^{(1)}|U_{1}^{(1)})\Big)+\tau_{2}\Big(I(U_{3}^{(2)};Y_{2}^{(2)})+I(X_{3}^{(2)};Y_{1}^{(2)}|U_{3}^{(2)})\Big)+\\
&+\tau_{3}I(X_{1}^{(3)}X_{3}^{(3)};Y_{2}^{(3)})+\tau_{5}I(X_{3}^{(5)};Y_{1}^{(5)}|X_{2}^{(5)})+\tau_{6}I(X_{1}^{(6)};Y_{3}^{(6)}|X_{2}^{(6)})
\end{align*}
with $0<\tau_{l}$ and $\sum_{l=1}^{6}\tau_{l}\leq1$, for some joint probability distributions of the form
\begin{align*}
P(u_{1}^{(1)} x_{1}^{(1)} y_{2}^{(1)} y_{3}^{(1)})&=P(u_{1}^{(1)})P(x_{1}^{(1)}|u_{1}^{(1)})P(y_{2}^{(1)} y_{3}^{(1)}|x_{1}^{(1)})\\
P(u_{3}^{(2)} x_{3}^{(2)} y_{2}^{(2)} y_{1}^{(2)})&=P(u_{3}^{(2)})P(x_{3}^{(2)}|u_{3}^{(2)})P(y_{1}^{(2)} y_{2}^{(2)}|x_{3}^{(2)})\\
P(x_{1}^{(3)} x_{3}^{(3)} y_{2}^{(3)})&=P(x_{1}^{(3)})P(x_{3}^{(3)})P(y_{2}^{(3)}|x_{1}^{(3)} x_{3}^{(3)})\\
P(x_{2}^{(4)} y_{1}^{(4)} y_{3}^{(4)})&=P(x_{2}^{(4)})P(y_{1}^{(4)} y_{3}^{(4)}|x_{2}^{(4)})\\
P(x_{2}^{(5)} x_{3}^{(5)} y_{1}^{(5)})&=P(x_{2}^{(5)})P(x_{3}^{(5)}|x_{2}^{(5)})P(y_{1}^{(5)}|x_{2}^{(5)} x_{3}^{(5)})\\
P(x_{1}^{(6)} x_{2}^{(6)} y_{3}^{(6)})&=P(x_{2}^{(6)})P(x_{1}^{(6)}|x_{2}^{(6)})P(y_{3}^{(6)}|x_{1}^{(6)} x_{2}^{(6)})
\end{align*}
are achievable with a 6P scheme and a partial-decode-and-forward (PDF) strategy.
\end{theorem}
\begin{proof}
For the proof it is assumed that the transmission is performed within $n\geq6$ channel uses and six subsequent phases, each with an individual network state, i.e., $L=6$. Phase $l$ features $n_l\geq1$ transmission slots and $\sum_{l=1}^{6}n_l=n$. If $n$ grows each $n_l$ is assumed to grow at the same speed. For large $n$, $\frac{n_l}{n}\to\tau_l>0$. The codebooks are labeled by the indices
\begin{align}
w_m=1,\ldots,2^{nR_m}
\end{align}
with $m=1,\ldots,14$ and $\hat{w}_m(i)$ denotes the estimate of the $m$-th index at node $i$.\hfill\\
\subsubsection{Codebook Construction}
Determine the input distributions $P_{X_1^{(1)}U_1^{(1)}}(\cdot)$, $P_{X_3^{(2)}U_3^{(2)}}(\cdot)$, $P_{X_1^{(3)}}(\cdot)$, $P_{X_3^{(3)}}(\cdot)$, $P_{X_2^{(4)}}(\cdot)$, $P_{X_3^{(5)}X_2^{(5)}}(\cdot)$ and $P_{X_1^{(6)}X_2^{(6)}}(\cdot)$.
Generate $2^{n(R_1+R_2+R_3)}$ $n_1$-sequences $u_{1}^{n_1}(w_{1},w_{2},w_{3})$ by choosing each $u_{1,k}^{(1)}$ independently according to the marginal $P_{U_1^{(1)}}(\cdot)$. 
For each $(w_{1},w_{2},w_{3})$, generate $2^{nR_4}$ $n_1$-sequences $x_{1}^{n_1}(w_{1},w_{2},w_{3},w_{4})$, choosing each $x_{1,k}^{(1)}$ independently according to $P_{X_1^{(1)}|U_1^{(1)}}(\cdot|u_{1,k}^{(1)}(w_{1},w_{2},w_{3}))$. 
Generate $2^{n(R_8+R_9+R_{10})}$ $n_2$-sequences $u_{3}^{n_2}(w_{8},w_{9},w_{10})$, choosing each $u_{3,k}^{(2)}$ independently according to the marginal $P_{U_3^{(2)}}(\cdot)$.
For each $(w_{8},w_{9},w_{10})$, generate $2^{nR_{11}}$ $n_2$-sequences $x_{3}^{n_2}(w_{8},w_{9},w_{10},w_{11})$, choosing each $x_{3,k}^{(2)}$ independently according to $P_{X_3^{(2)}|U_3^{(2)}}(\cdot|u_{3,k}^{(2)}(w_{8},w_{9},w_{10}))$. 
Generate $2^{n(R_5+R_6)}$ $n_3$-sequences $x_{1}^{n_3}(w_{5},w_{6})$, choosing each $x_{1,k}^{(3)}$ independently according to $P_{X_1^{(3)}}(\cdot)$.
Generate $2^{n(R_{12}+R_{13})}$ $n_3$-sequences $x_{3}^{n_3}(w_{12},w_{13})$, choosing each $x_{3,k}^{(3)}$ independently according to $P_{X_3^{(3)}}(\cdot)$.
Generate  $2^{n(R_{1}+R_{5}+R_{8}+R_{12})}$ $n_4$-sequences $x_{2}^{n_4}(w_{1},w_{5},w_{8},w_{12})$, choosing each $x_{2,k}^{(4)}$ independently according to $P_{X_2^{(4)}}(\cdot)$.
Generate  $2^{n(R_{9}+R_{13})}$ $n_5$-sequences $x_{2}^{n_5}(w_{9},w_{13})$, choosing each $x_{2,k}^{(5)}$ independently according to $P_{X_2^{(5)}}(\cdot)$.
For each $x_{2}^{n_5}(w_{9},w_{13})$ generate $2^{nR_{14}}$ $n_5$-sequences $x_{3}^{n_5}(w_{9},w_{13},w_{14})$, choosing each $x_{3,k}^{(5)}$ independently according to $P_{X_3^{(5)}|X_2^{(5)}}(\cdot|x_{2,k}^{(5)}(w_{9},w_{13}))$. 
Generate $2^{n(R_{2}+R_{6})}$ $n_6$-sequences $x_{2}^{n_6}(w_{2},w_{6})$, choosing each $x_{2,k}^{(6)}$ independently according to $P_{X_2^{(6)}}(\cdot)$.
For each $x_{2}^{n_6}(w_{2},w_{6})$ generate $2^{nR_{7}}$ $n_6$-sequences $x_{1}^{n_6}(w_{2},w_{6},w_{7})$, choosing each $x_{1,k}^{(6)}$ independently according to $P_{X_1^{(6)}|X_2^{(6)}}(\cdot|x_{2,k}^{(6)}(w_{2},w_{6}))$.
%
%
\subsubsection{Input at node 1}
The message $w_{1\text{-}3}$ is reindexed by 
\begin{align}
(w_{1},w_{2},w_{3},w_{4},w_{5},w_{6},w_{7}).
\end{align}
In phase 1, 3 and 6 node 1 transmits the sequences
\begin{align}
&x_{1}^{n_1}(w_{1},w_{2},w_{3},w_{4})\notag\\
&x_{1}^{n_3}(w_{5},w_{6})\notag\\
&x_{1}^{n_6}(w_{2},w_{6},w_{7}).
\end{align}
%
%
\subsubsection{Input at node $3$}
The message $w_{3\text{-}1}$ is reindexed by 
\begin{align}
(w_{8},w_{9},w_{10},w_{11},w_{12},w_{13},w_{14}).
\end{align}
In phase 2, 3, and 5 node 3 transmits the sequences
\begin{align}
&x_{3}^{n_2}(w_{8},w_{9},w_{10},w_{11})\notag\\
&x_{3}^{n_3}(w_{12},w_{13})\notag\\
&x_{3}^{n_5}(w_{9},w_{13},w_{14}).
\end{align}
%
%
\subsubsection{Output at node 2}
The sequences $y_{2}^{n_1}$, $y_{2}^{n_2}$ and $y_{2}^{n_3}$ are observed in phase 1,2, and 3. After that node 2 tries to find $(\tilde{w}_{1},\tilde{w}_{2},\tilde{w}_{3})$ such that
\begin{align}
\Big( u_{1}^{n_1}(\tilde{w}_{1},\tilde{w}_{2},\tilde{w}_{3}),y_{2}^{n_1}\Big)\in T_{\epsilon}^{n_1}(P_{U_1^{(1)}Y_2^{(1)}}).
\end{align}
If there is none or more than one such $(\tilde{w}_{1},\tilde{w}_{2},\tilde{w}_{3})$ an error is declared. Otherwise, the found $(\tilde{w}_{1},\tilde{w}_{2},\tilde{w}_{3})$ is the estimate $(\hat{w}_{1}(2),\hat{w}_{2}(2),\hat{w}_{3}(2))$. 
Then node 2 tries to find $(\tilde{w}_{8},\tilde{w}_{9},\tilde{w}_{10})$ such that
\begin{align}
\Big(u_{3}^{n_2}(\tilde{w}_{8},\tilde{w}_{9},\tilde{w}_{10}),y_{2}^{n_2} \Big)\in T_{\epsilon}^{n_2}(P_{U_3^{(2)}Y_2^{(2)}}).
\end{align}
If there is none or more than one such $(\tilde{w}_{8},\tilde{w}_{9},\tilde{w}_{10})$ an error is declared. Otherwise, the found $(\tilde{w}_{8},\tilde{w}_{9},\tilde{w}_{10})$ is the estimate $(\hat{w}_{8}(2),\hat{w}_{9}(2),\hat{w}_{10}(2))$. 
Then node 2 tries to find $(\tilde{w}_{5},\tilde{w}_{6},\tilde{w}_{12},\tilde{w}_{13})$ such that
\begin{align}
\Big(x_{1}^{n_3}(\tilde{w}_{5},\tilde{w}_{6}),x_{3}^{n_3}(\tilde{w}_{12},\tilde{w}_{13}),y_{2}^{n_3} \Big)\in T_{\epsilon}^{n_3}(P_{X_1^{(3)}X_3^{(3)}Y_2^{(3)}}).
\end{align}
If there is none or more than one such $(\tilde{w}_{5},\tilde{w}_{6},\tilde{w}_{12},\tilde{w}_{13})$ an error is declared. Otherwise, the found $(\tilde{w}_{5},\tilde{w}_{6},\tilde{w}_{12},\tilde{w}_{13})$ is the estimate $(\hat{w}_{5}(2),\hat{w}_{6}(2),\hat{w}_{12}(2),\hat{w}_{13}(2))$.
%
%
\subsubsection{Input at node 2}
In phase 4, 5 and 6 node 2 sends the sequences 
\begin{align}
&x_{2}^{n_4}(\hat{w}_{1}(2),\hat{w}_{5}(2),\hat{w}_{8}(2),\hat{w}_{12}(2))\notag\\
&x_{2}^{n_5}(\hat{w}_{9}(2),\hat{w}_{13}(2))\notag\\
&x_{2}^{n_6}(\hat{w}_{2}(2),\hat{w}_{6}(2)).
\end{align}
%
%
\subsubsection{Output at node 1}
In phase 2, 4 and 5 the sequences $y_{1}^{n_2}$, $y_{1}^{n_4}$ and $y_{1}^{n_5}$ are observed. After that node 1 tries to find $(\tilde{w}_{8},\tilde{w}_{12})$ such that
\begin{align}
\Big(x_{2}^{n_4}(w_{1},w_{5},\tilde{w}_{8},\tilde{w}_{12}),y_{1}^{n_4} \Big)\in T_{\epsilon}^{n_4}(P_{X_2^{(4)}Y_1^{(4)}}).
\end{align}
If there is none or more than one such $(\tilde{w}_{8},\tilde{w}_{12})$ an error is declared. Otherwise, the found $(\tilde{w}_{8},\tilde{w}_{12})$ is the estimate $(\hat{w}_{8}(1),\hat{w}_{12}(1))$.
Then node 1 tries to find $(\tilde{w}_{9},\tilde{w}_{13})$ such that
\begin{align}
\Big(x_{2}^{n_5}(\tilde{w}_{9},\tilde{w}_{13}),y_{1}^{n_5} \Big)\in T_{\epsilon}^{n_5}(P_{X_2^{(5)}Y_1^{(5)}}).
\end{align}
If there is none or more than one such $(\tilde{w}_{9},\tilde{w}_{13})$ an error is declared. Otherwise, the found $(\tilde{w}_{9},\tilde{w}_{13})$ is the estimate $(\hat{w}_{9}(1),\hat{w}_{13}(1))$.
Then node 1 tries to find a $\tilde{w}_{14}$ such that
\begin{align}
\Big(x_{3}^{n_5}(\hat{w}_{9}(1),\hat{w}_{13}(1),\tilde{w}_{14}),x_{2}^{n_5}(\hat{w}_{9}(1),\hat{w}_{13}(1)),y_{1}^{n_5} \Big)\in T_{\epsilon}^{n_5}(P_{X_3^{(5)}X_2^{(5)}Y_1^{(5)}}).
\end{align}
If there is none or more than one such $\tilde{w}_{14}$ an error is declared. Otherwise, the found $\tilde{w}_{14}$ is the estimate $\hat{w}_{14}(1)$.
Then node 1 tries to find a $\tilde{w}_{10}$ such that
\begin{align}
\Big(u_{3}^{n_2}(\hat{w}_{8}(1),\hat{w}_{9}(1),\tilde{w}_{10}),y_{1}^{n_2} \Big)\in T_{\epsilon}^{n_2}(P_{U_3^{(2)}Y_1^{(2)}}).
\end{align}
If there is none or more than one such $\tilde{w}_{10}$ an error is declared. Otherwise, the found $\tilde{w}_{10}$ is the estimate $\hat{w}_{10}(1)$. 
Finally node 1 tries to find a $\tilde{w}_{11}$ such that
\begin{align}
\Big(u_{3}^{n_2}(\hat{w}_{8}(1),\hat{w}_{9}(1),\hat{w}_{10}(1)),x_{3}^{n_2}(\hat{w}_{8}(1),\hat{w}_{9}(1),\hat{w}_{10}(1),\tilde{w}_{11}),y_{1}^{n_2} \Big)\in T_{\epsilon}^{n_2}(P_{U_3^{(2)}X_3^{(2)}Y_1^{(2)}}).
\end{align}
If there is none or more than one such $\tilde{w}_{11}$ an error is declared. Otherwise, the found $\tilde{w}_{11}$ is the estimate $\hat{w}_{11}(1)$. The message estimate $\hat{w}_{3\text{-}1}(1)$ is found by reindexing 
\begin{align}
(\hat{w}_{8}(1),\hat{w}_{9}(1),\hat{w}_{10}(1),\hat{w}_{11}(1),\hat{w}_{12}(1),\hat{w}_{13}(1),\hat{w}_{14}(1)).
\end{align}
%
%
\subsubsection{Output at node 3}
In phase 1, 4 and 6 the sequences $y_{3}^{n_1}$, $y_{3}^{n_4}$ and $y_{3}^{n_6}$ are observed. After that node 3 tries to find $(\tilde{w}_{1},\tilde{w}_{5})$ such that
\begin{align}
\Big(x_{2}^{n_4}(\tilde{w}_{1},\tilde{w}_{5},w_{8},w_{12}),y_{3}^{n_4} \Big)\in T_{\epsilon}^{n_4}(P_{X_2^{(4)}Y_3^{(4)}}).
\end{align}
If there is none or more than one such $(\tilde{w}_{1},\tilde{w}_{5})$ an error is declared. Otherwise, the found $(\tilde{w}_{1},\tilde{w}_{5})$ is the estimate $(\hat{w}_{1}(3),\hat{w}_{5}(3))$.
Then node 3 tries to find $(\tilde{w}_{2},\tilde{w}_{6})$ such that
\begin{align}
\Big(x_{2}^{n_6}(\tilde{w}_{2},\tilde{w}_{6}),y_{3}^{n_6} \Big)\in T_{\epsilon}^{n_6}(P_{X_2^{(6)}Y_3^{(6)}}).
\end{align}
If there is none or more than one such $(\tilde{w}_{2},\tilde{w}_{6})$ an error is declared. Otherwise, the found $(\tilde{w}_{2},\tilde{w}_{6})$ is the estimate $(\hat{w}_{2}(3),\hat{w}_{6}(3))$.
Then node 3 tries to find a $\tilde{w}_{7}$ such that
\begin{align}
\Big(x_{1}^{n_6}(\hat{w}_{2}(3),\hat{w}_{6}(3),\tilde{w}_{7}),x_{2}^{n_6}(\hat{w}_{2}(3),\hat{w}_{6}(3)),y_{3}^{n_6} \Big) \in T_{\epsilon}^{n_6}(P_{X_1^{(6)}X_2^{(6)}Y_3^{(6)}}).
\end{align}
If there is none or more than one such $\tilde{w}_{7}$ an error is declared. Otherwise, the found $\tilde{w}_{7}$ is the estimate $\hat{w}_{7}(3)$.
Then node 3 tries to find a $\tilde{w}_{3}$ such that
\begin{align}
\Big(u_{1}^{n_1}(\hat{w}_{1}(3),\hat{w}_{2}(3),\tilde{w}_{3}),y_{3}^{n_1} \Big)\in T_{\epsilon}^{n_1}(P_{U_1^{(1)}Y_3^{(1)}}).
\end{align}
If there is none or more than one such $\tilde{w}_{3}$ an error is declared. Otherwise, the found $\tilde{w}_{3}$ is the estimate $\hat{w}_{3}(3)$. 
Finally node 3 tries to find a $\tilde{w}_{4}$ such that
\begin{align}
\Big(u_{1}^{n_1}(\hat{w}_{1}(3),\hat{w}_{2}(3),\hat{w}_{3}(3)),x_{1}^{n_1}(\hat{w}_{1}(3),\hat{w}_{2}(3),\hat{w}_{3}(3),\tilde{w}_{4}),y_{3}^{n_1} \Big)\in T_{\epsilon}^{n_1}(P_{U_{1}^{(1)}X_{1}^{(1)}Y_3^{(1)}}).
\end{align}
If there is none or more than one such $\tilde{w}_{4}$ an error is declared. Otherwise, the found $\tilde{w}_{4}$ is the estimate $\hat{w}_{4}(3)$. The message estimate $\hat{w}_{1\text{-}3}(3)$ is found by reindexing 
\begin{align}
(\hat{w}_{1}(3),\hat{w}_{2}(3),\hat{w}_{3}(3),\hat{w}_{4}(3),\hat{w}_{5}(3),\hat{w}_{6}(3),\hat{w}_{7}(3)).
\end{align}
%
%
\subsubsection{Error Analysis}
The average error probability $\bar{P}_{e}$ can be upper bounded by the sum of the error probabilities of each individual decoding step calculated under the assumption that all prior decoding steps in the network have been performed error-free. According to the properties of \emph{$\epsilon$-letter typical sequences} \cite{Kra07}, the probabilities of the individual errors can be made arbitrarily small by choosing $n$ sufficiently large while the rate constraints
\begin{align}
R_1+R_2+R_3&<\tau_1I(U_{1}^{(1)};Y_{2}^{(1)})-\delta(\epsilon)\label{aconst:1}\\
R_8+R_9+R_{10}&<\tau_2I(U_{3}^{(2)};Y_{2}^{(2)})-\delta(\epsilon)\label{aconst:2}\\
R_5+R_6&<\tau_3I(X_{1}^{(3)};Y_{2}^{(3)}|X_{3}^{(3)})-\delta(\epsilon)\label{aconst:3}\\
R_{12}+R_{13}&<\tau_3I(X_{3}^{(3)};Y_{2}^{(3)}|X_{1}^{(3)})-\delta(\epsilon)\label{aconst:4}\\
R_5+R_6+R_{12}+R_{13}&<\tau_3I(X_{1}^{(3)}X_{3}^{(3)};Y_{2}^{(3)})-\delta(\epsilon)\label{aconst:5}
\end{align}
at node 2,
\begin{align}
R_{8}+R_{12}&<\tau_4I(X_{2}^{(4)};Y_{1}^{(4)})-\delta(\epsilon)\label{aconst:6}\\
R_{9}+R_{13}&<\tau_5I(X_{2}^{(5)};Y_{1}^{(5)})-\delta(\epsilon)\label{aconst:7}\\
R_{14}&<\tau_5I(X_{3}^{(5)};Y_{1}^{(5)}|X_{2}^{(5)})-\delta(\epsilon)\label{aconst:8}\\
R_{10}&<\tau_2I(U_{3}^{(2)};Y_{1}^{(2)})-\delta(\epsilon)\label{aconst:9}\\
R_{11}&<\tau_2I(X_{3}^{(2)};Y_{1}^{(2)}|U_{3}^{(2)})-\delta(\epsilon)\label{aconst:10}
\end{align}
at node 1 and
\begin{align}
R_{1}+R_{5}&<\tau_4I(X_{2}^{(4)};Y_{3}^{(4)})-\delta(\epsilon)\label{aconst:11}\\
R_{2}+R_{6}&<\tau_6I(X_{2}^{(6)};Y_{3}^{(6)})-\delta(\epsilon)\label{aconst:12}\\
R_{7}&<\tau_6I(X_{1}^{(6)};Y_{3}^{(6)}|X_{2}^{(6)})-\delta(\epsilon)\label{aconst:13}\\
R_{3}&<\tau_1I(U_{1}^{(1)};Y_{3}^{(1)})-\delta(\epsilon)\label{aconst:14}\\
R_{4}&<\tau_1I(X_{1}^{(1)};Y_{3}^{(1)}|U_{1}^{(2)})-\delta(\epsilon)\label{aconst:15}
\end{align}
at node 3 are satisfied, while $\delta(\epsilon)$ can be made small by choosing $\epsilon$ small. In order to proof Theorem \ref{prop:AR_6P_PDF} we will show that if the constraints
\begin{align}
R_{1\text{-}3}&<\tau_1I(U_{1}^{(1)};Y_{2}^{(1)})+\tau_1I(X_{1}^{(1)};Y_{3}^{(1)}|U_{1}^{(1)})+\tau_3I(X_{1}^{(3)};Y_{2}^{(3)}|X_{3}^{(3)})+\tau_6I(X_{1}^{(6)};Y_{3}^{(6)}|X_{2}^{(6)})-7\delta(\epsilon)\notag\\
R_{1\text{-}3}&<\tau_1I(U_{1}^{(1)};Y_{3}^{(1)})+\tau_1I(X_{1}^{(1)};Y_{3}^{(1)}|U_{1}^{(1)})+\tau_4I(X_{2}^{(4)};Y_{3}^{(4)})+\tau_6I(X_{2}^{(6)};Y_{3}^{(6)})+\notag\\&+\tau_6I(X_{1}^{(6)};Y_{3}^{(6)}|X_{2}^{(6)})-9\delta(\epsilon)\notag\\
R_{3\text{-}1}&<\tau_2I(U_{3}^{(2)};Y_{2}^{(2)})+\tau_2I(X_{3}^{(2)};Y_{1}^{(2)}|U_{3}^{(2)})+\tau_3I(X_{3}^{(3)};Y_{2}^{(3)}|X_{1}^{(3)})+\tau_5I(X_{3}^{(5)};Y_{1}^{(5)}|X_{2}^{(5)})-7\delta(\epsilon)\notag\\
R_{3\text{-}1}&<\tau_2I(U_{3}^{(2)};Y_{1}^{(2)})+\tau_2I(X_{3}^{(2)};Y_{1}^{(2)}|U_{3}^{(2)})+\tau_4I(X_{2}^{(4)};Y_{1}^{(4)})+\tau_5I(X_{2}^{(5)};Y_{1}^{(5)})+\notag\\&+\tau_5I(X_{3}^{(5)};Y_{1}^{(5)}|X_{2}^{(5)})-9\delta(\epsilon)\notag
\end{align}
subject to
\begin{align}
R_{1\text{-}3}+R_{3\text{-}1}&<\tau_1I(U_{1}^{(1)};Y_{2}^{(1)})+\tau_1I(X_{1}^{(1)};Y_{3}^{(1)}|U_{1}^{(1)})+\tau_2I(U_{3}^{(2)};Y_{2}^{(2)})+\tau_2I(X_{3}^{(2)};Y_{1}^{(2)}|U_{3}^{(2)})+\notag\\
&+\tau_3I(X_{1}^{(3)}X_{3}^{(3)};Y_{2}^{(3)})+\tau_5I(X_{3}^{(5)};Y_{1}^{(5)}|X_{2}^{(5)})+\tau_6I(X_{1}^{(6)};Y_{3}^{(6)}|X_{2}^{(6)})-13\delta(\epsilon)\label{aconst:sys}
\end{align}
are satisfied it is possible to construct a code with
\begin{align}
R_{1\text{-}3}&=R_{1}+R_{2}+R_{3}+R_{4}+R_{5}+R_{6}+R_{7}\notag\\
R_{3\text{-}1}&=R_{8}+R_{9}+R_{10}+R_{11}+R_{12}+R_{13}+R_{14}
\end{align}
that satisfies the constraints (\ref{aconst:1}) to (\ref{aconst:15}): Set
\begin{align}
R_{14}&=\tau_5I(X_{3}^{(5)};Y_{1}^{(5)}|X_{2}^{(5)})-2\delta(\epsilon)\notag\\
R_{10}&=\tau_2I(U_{3}^{(2)};Y_{1}^{(2)})-2\delta(\epsilon)\notag\\
R_{11}&=\tau_2I(X_{3}^{(2)};Y_{1}^{(2)}|U_{3}^{(2)})-2\delta(\epsilon)\notag\\
R_{7}&=\tau_6I(X_{1}^{(6)};Y_{3}^{(6)}|X_{2}^{(6)})-2\delta(\epsilon)\notag\\
R_{3}&=\tau_1I(U_{1}^{(1)};Y_{3}^{(1)})-2\delta(\epsilon)\notag\\
R_{4}&=\tau_1I(X_{1}^{(1)};Y_{3}^{(1)}|U_{1}^{(1)})-2\delta(\epsilon).
\end{align}
This satisfies constraints (\ref{aconst:8}), (\ref{aconst:9}), (\ref{aconst:10}), (\ref{aconst:13}), (\ref{aconst:14}) and (\ref{aconst:15}). The constraints (\ref{aconst:1}) and (\ref{aconst:2}) change to
\begin{align}
R_1+R_2&<\tau_1I(U_{1}^{(1)};Y_{2}^{(1)})-\tau_1I(U_{1}^{(1)};Y_{3}^{(1)})+\delta(\epsilon)\label{aconst:1mod}\\
R_8+R_9&<\tau_2I(U_{3}^{(2)};Y_{2}^{(2)})-\tau_2I(U_{3}^{(2)};Y_{1}^{(2)})+\delta(\epsilon)\label{aconst:2mod}
\end{align}
and (\ref{aconst:sys}) can be written as
\begin{align}
R_{1}+R_{2}+R_{5}+R_{6}&<\tau_1I(U_{1}^{(1)};Y_{2}^{(1)})-\tau_1I(U_{1}^{(1)};Y_{3}^{(1)})+\tau_3I(X_{1}^{(3)};Y_{2}^{(3)}|X_{3}^{(3)})-\delta(\epsilon)\notag\\
R_{1}+R_{2}+R_{5}+R_{6}&<\tau_4I(X_{2}^{(4)};Y_{3}^{(4)})+\tau_6I(X_{2}^{(6)};Y_{3}^{(6)})-3\delta(\epsilon)\notag\\
R_{8}+R_{9}+R_{12}+R_{13}&<\tau_2I(U_{3}^{(2)};Y_{2}^{(2)})-\tau_2I(U_{3}^{(2)};Y_{1}^{(2)})+\tau_3I(X_{3}^{(3)};Y_{2}^{(3)}|X_{1}^{(3)})-\delta(\epsilon)\notag\\
R_{8}+R_{9}+R_{12}+R_{13}&<\tau_4I(X_{2}^{(4)};Y_{1}^{(4)})+\tau_5I(X_{2}^{(5)};Y_{1}^{(5)})-3\delta(\epsilon)\notag
\end{align}
subject to
\begin{align}
R_{1}+R_{2}+R_{5}+R_{6}+R_{8}+R_{9}+R_{12}+R_{13}&<\tau_1I(U_{1}^{(1)};Y_{2}^{(1)})-\tau_1I(U_{1}^{(1)};Y_{3}^{(1)})+\tau_2I(U_{3}^{(2)};Y_{2}^{(2)})-\notag\\&-\tau_2I(U_{3}^{(2)};Y_{1}^{(2)})+\tau_3I(X_{1}^{(3)}X_{3}^{(3)};Y_{2}^{(3)})-\delta(\epsilon)\label{aconst:sysmod}.
\end{align}
Set
\begin{align}
R_1&=\tau_1I(U_{1}^{(1)};Y_{2}^{(1)})-\tau_1I(U_{1}^{(1)};Y_{3}^{(1)})-R_2\notag\\
R_8&=\tau_2I(U_{3}^{(2)};Y_{2}^{(2)})-\tau_2I(U_{3}^{(2)};Y_{1}^{(2)})-R_9.
\end{align}
This satisfies (\ref{aconst:1}) through (\ref{aconst:1mod}) and (\ref{aconst:2}) through (\ref{aconst:2mod}) while (\ref{aconst:6}) and (\ref{aconst:11}) change to 
\begin{align}
R_{5}&<\tau_1I(U_{1}^{(1)};Y_{3}^{(1)})-\tau_1I(U_{1}^{(1)};Y_{2}^{(1)})+\tau_4I(X_{2}^{(4)};Y_{3}^{(4)})+R_2-\delta(\epsilon)\label{aconst:6mod}\\
R_{12}&<\tau_2I(U_{3}^{(2)};Y_{1}^{(2)})-\tau_2I(U_{3}^{(2)};Y_{2}^{(2)})+\tau_4I(X_{2}^{(4)};Y_{1}^{(4)})+R_9-\delta(\epsilon)\label{aconst:11mod}.
\end{align}
The equations (\ref{aconst:sysmod}/\ref{aconst:sys}) change to
\begin{align}
R_{5}+R_{6}&<\tau_3I(X_{1}^{(3)};Y_{2}^{(3)}|X_{3}^{(3)})-\delta(\epsilon)\notag\\
R_{5}+R_{6}&<\tau_1I(U_{1}^{(1)};Y_{3}^{(1)})-\tau_1I(U_{1}^{(1)};Y_{2}^{(1)})+\tau_4I(X_{2}^{(4)};Y_{3}^{(4)})+\tau_6I(X_{2}^{(6)};Y_{3}^{(6)})-3\delta(\epsilon)\notag\\
R_{12}+R_{13}&<\tau_3I(X_{3}^{(3)};Y_{2}^{(3)}|X_{1}^{(3)})-\delta(\epsilon)\notag\\
R_{12}+R_{13}&<\tau_2I(U_{3}^{(2)};Y_{1}^{(2)})-\tau_2I(U_{3}^{(2)};Y_{2}^{(2)})+\tau_4I(X_{2}^{(4)};Y_{1}^{(4)})+\tau_5I(X_{2}^{(5)};Y_{1}^{(5)})-3\delta(\epsilon)\notag
\end{align}
subject to
\begin{align}
R_{5}+R_{6}+R_{12}+R_{13}&<\tau_3I(X_{1}^{(3)}X_{3}^{(3)};Y_{2}^{(3)})-\delta(\epsilon)\label{aconst:sysmod2}.
\end{align}
Set
\begin{align}
R_{9}&=\tau_5I(X_{2}^{(5)};Y_{1}^{(5)})-R_{13}-2\delta(\epsilon)\\
R_{2}&=\tau_6I(X_{2}^{(6)};Y_{3}^{(6)})-R_{6}-2\delta(\epsilon).
\end{align}
This satisfies (\ref{aconst:7}) and (\ref{aconst:12}) while (\ref{aconst:6mod}/\ref{aconst:6}) and (\ref{aconst:11mod}/\ref{aconst:11}) are
\begin{align}
R_{5}+R_{6}&<\tau_1I(U_{1}^{(1)};Y_{3}^{(1)})-\tau_1I(U_{1}^{(1)};Y_{2}^{(1)})+\tau_4I(X_{2}^{(4)};Y_{3}^{(4)})+\tau_6I(X_{2}^{(6)};Y_{3}^{(6)})-3\delta(\epsilon)\label{aconst:6mod2}\\
R_{12}+R_{13}&<\tau_2I(U_{3}^{(2)};Y_{1}^{(2)})-\tau_2I(U_{3}^{(2)};Y_{2}^{(2)})+\tau_4I(X_{2}^{(4)};Y_{1}^{(4)})+\tau_5I(X_{2}^{(5)};Y_{1}^{(5)})-3\delta(\epsilon)\label{aconst:11mod2}.
\end{align}
It is now evident that (\ref{aconst:sysmod2}/\ref{aconst:sysmod}/\ref{aconst:sys}) also satisfies the remaining constraints (\ref{aconst:3}),  (\ref{aconst:4}) and (\ref{aconst:5}) as well as (\ref{aconst:6}) through (\ref{aconst:6mod2}/\ref{aconst:6mod}) and (\ref{aconst:11}) through (\ref{aconst:11mod2}/\ref{aconst:11mod}). Achievability of Theorem \ref{prop:AR_6P_PDF} follows by making $\epsilon$ in (\ref{aconst:sys}) small and using the fact that
\begin{align}
U_{1}^{(1)}-X_{1}^{(1)}-Y_{3}^{(1)}\notag\\
U_{3}^{(2)}-X_{3}^{(2)}-Y_{1}^{(2)}
\end{align}
form Markov chains.
\end{proof}
A simpler second bound follows directly as a special case of Theorem \ref{prop:AR_6P_PDF}. Note that this strategy requires full decoding of the dialog messages at the relay in the first and second phase.
\begin{theorem}\label{prop:AR_6P_DF}
All rate pairs of the half-duplex two-way relay channel that satisfy
\begin{align*}
R_{1\text{-}3} &\leq \tau_{1}I(X_{1}^{(1)};Y_{2}^{(1)}) + \tau_{3}I(X_{1}^{(3)};Y_{2}^{(3)}|X_{3}^{(3)})+ \tau_{6}I(X_{1}^{(6)};Y_{3}^{(6)}|X_{2}^{(6)})\\
R_{1\text{-}3} &\leq \tau_{1} I(X_{1}^{(1)};Y_{3}^{(1)}) + \tau_{4} I(X_{2}^{(4)};Y_{3}^{(4)})+ \tau_{6} I(X_{1}^{(6)}X_{2}^{(6)};Y_{3}^{(6)})\\
R_{3\text{-}1} &\leq \tau_{2}I(X_{3}^{(2)};Y_{2}^{(2)}) + \tau_{3}I(X_{3}^{(3)};Y_{2}^{(3)}|X_{1}^{(3)})+ \tau_{5}I(X_{3}^{(5)};Y_{1}^{(5)}|X_{2}^{(5)})\\
R_{3\text{-}1} &\leq \tau_{2} I(X_{3}^{(2)};Y_{1}^{(2)}) + \tau_{4} I(X_{2}^{(4)};Y_{1}^{(4)})+\tau_{5} I(X_{2}^{(5)}X_{3}^{(5)};Y_{1}^{(5)})
\end{align*}
subject to
\begin{align*}
R_{1\text{-}3}+R_{3\text{-}1} &\leq \tau_{1}I(X_{1}^{(1)};Y_{2}^{(1)})+\tau_{2}I(X_{3}^{(2)};Y_{2}^{(2)})+\tau_{3}I(X_{1}^{(3)}X_{3}^{(3)};Y_{2}^{(3)})+\notag\\
&+\tau_{5}I(X_{3}^{(5)};Y_{1}^{(5)}|X_{2}^{(5)})+\tau_{6}I(X_{1}^{(6)};Y_{3}^{(6)}|X_{2}^{(6)})
\end{align*}
with $0<\tau_{l}$ and $\sum_{l=1}^{6}\tau_{l}\leq1$, for some joint probability distributions of the form
\begin{align*}
P(x_{1}^{(1)} y_{2}^{(1)} y_{3}^{(1)})&=P(x_{1}^{(1)})P(y_{2}^{(1)} y_{3}^{(1)}|x_{1}^{(1)})\\
P(x_{3}^{(2)} y_{2}^{(2)} y_{1}^{(2)})&=P(x_{3}^{(2)})P(y_{1}^{(2)} y_{2}^{(2)}|x_{3}^{(2)})\\
P(x_{1}^{(3)} x_{3}^{(3)} y_{2}^{(3)})&=P(x_{1}^{(3)})P(x_{3}^{(3)})P(y_{2}^{(3)}|x_{1}^{(3)} x_{3}^{(3)})\\
P(x_{2}^{(4)} y_{1}^{(4)} y_{3}^{(4)})&=P(x_{2}^{(4)})P(y_{1}^{(4)} y_{3}^{(4)}|x_{2}^{(4)})\\
P(x_{2}^{(5)} x_{3}^{(5)} y_{1}^{(5)})&=P(x_{2}^{(5)})P(x_{3}^{(5)}|x_{2}^{(5)})P(y_{1}^{(5)}|x_{2}^{(5)} x_{3}^{(5)})\\
P(x_{1}^{(6)} x_{2}^{(6)} y_{3}^{(6)})&=P(x_{2}^{(6)})P(x_{1}^{(6)}|x_{2}^{(6)})P(y_{3}^{(6)}|x_{1}^{(6)} x_{2}^{(6)})
\end{align*}
are achievable with a 6P scheme and a decode-and-forward (DF) strategy.
\end{theorem}
\begin{proof}
In Theorem \ref{prop:AR_6P_PDF} set $U_{1}^{(1)}=X_{1}^{(1)}$ and $U_{3}^{(2)}=X_{3}^{(2)}$ with the effect that
\begin{align}
I(X_{1}^{(1)};Y_{3}^{(1)}|U_{1}^{(1)})=I(X_{3}^{(2)};Y_{1}^{(2)}|U_{3}^{(2)})=0.
\end{align}
\end{proof}
Note that Theorem \ref{prop:UB_6P}, \ref{prop:AR_6P_PDF} and \ref{prop:AR_6P_DF} provide rate bounds under a fixed set of input distributions. Such bounds can be extended to form capacity bounds by taking the union of rate regions over all input distributions and time-allocations $P_S(l)=\tau_l, \forall l$ (after adding a time-sharing variable $Q$).
\section{Wireline Communication}
Returning to the initial toy example let us consider the special case of orthogonal links. For such a case the channel can be defined by
\begin{align}
&(\mathcal{X}_{12}\times\mathcal{X}_{13}\times\mathcal{X}_{21}\times\mathcal{X}_{23}\times\mathcal{X}_{31}\times\mathcal{X}_{32},\notag\\
&P(y_{12}y_{13}y_{21}y_{23}y_{31}y_{32}|x_{12}x_{13} x_{21}x_{23} x_{31} x_{32} s),\notag\\
&\mathcal{Y}_{12}\times\mathcal{Y}_{13}\times\mathcal{Y}_{21}\times\mathcal{Y}_{23}\times\mathcal{Y}_{31}\times\mathcal{Y}_{32},\mathcal{S})
\end{align}
where $X_{ij}$ is the input at the $i$-th node for communication to the $j$-th node and $Y_{ij}$ is the channel output at the $i$-th node listening exclusively to the $j$-th node in the network. Orthogonality restricts the channel to factorize
\begin{align}
P(y_{12}y_{13}y_{21}y_{23}y_{31}y_{32}|x_{12}x_{13} x_{21}x_{23} x_{31} x_{32} s)=
&P(y_{12}|x_{12} x_{21} s)
P(y_{13}|x_{13} x_{31} s)
P(y_{21}|x_{12} x_{21} s)\cdot\notag\\
\cdot &P(y_{23}|x_{23} x_{32} s)
P(y_{31}|x_{13} x_{31} s)
P(y_{32}|x_{23} x_{32} s).
\end{align}
\begin{theorem}\label{prop:AR_6P_PDF_ORT}
All rate pairs of the half-duplex two-way relay channel with orthogonal links that satisfy
\begin{align*}
R_{1\text{-}3} &\leq \tau_{1}\Big(I(X_{12}^{(1)};Y_{21}^{(1)})+I(X_{13}^{(1)};Y_{31}^{(1)})\Big) + \tau_{3}I(X_{12}^{(3)};Y_{21}^{(3)})+ \tau_{6}I(X_{13}^{(6)};Y_{31}^{(6)})\\
R_{1\text{-}3} &\leq \tau_{1} I(X_{13}^{(1)};Y_{31}^{(1)}) + \tau_{4} I(X_{23}^{(4)};Y_{32}^{(4)})+ \tau_{6}\Big( I(X_{13}^{(6)};Y_{31}^{(6)}) + I(X_{23}^{(6)};Y_{32}^{(6)})\Big)\\
R_{3\text{-}1} &\leq \tau_{2}\Big(I(X_{32}^{(2)};Y_{23}^{(2)})+I(X_{31}^{(2)};Y_{13}^{(2)})\Big) + \tau_{3}I(X_{32}^{(3)};Y_{23}^{(3)})+ \tau_{5}I(X_{31}^{(5)};Y_{12}^{(5)})\\
R_{3\text{-}1} &\leq \tau_{2} I(X_{31}^{(2)};Y_{13}^{(2)}) + \tau_{4} I(X_{21}^{(4)};Y_{12}^{(4)})+\tau_{5} \Big( I(X_{21}^{(5)};Y_{12}^{(5)}) +  I(X_{31}^{(5)};Y_{13}^{(5)}) \Big)
\end{align*}
with $0<\tau_{l}$ and $\sum_{l=1}^{6}\tau_{l}\leq1$, for some joint probability distributions of the form
\begin{align*}
P(x_{12}^{(1)} x_{13}^{(1)} y_{21}^{(1)} y_{31}^{(1)})&=P(x_{12}^{(1)})P(x_{13}^{(1)}|x_{12}^{(1)})P(y_{21}^{(1)} |x_{12}^{(1)})P( y_{31}^{(1)}|x_{13}^{(1)})\\
P(x_{31}^{(2)} x_{32}^{(2)} y_{13}^{(2)} y_{23}^{(2)})&=P(x_{31}^{(2)}|x_{32}^{(2)})P(x_{32}^{(2)})P(y_{13}^{(2)}|x_{31}^{(2)})P(y_{23}^{(2)}|x_{32}^{(2)})\\
P(x_{12}^{(3)} x_{32}^{(3)} y_{21}^{(3)} y_{23}^{(3)})&=P(x_{12}^{(3)})P(x_{32}^{(3)})P(y_{21}^{(3)}|x_{12}^{(3)} )P(y_{23}^{(3)}|x_{32}^{(3)} )\\
P(x_{21}^{(4)} x_{23}^{(4)} y_{12}^{(4)} y_{32}^{(4)})&=P(x_{21}^{(4)})P(x_{23}^{(4)})P(y_{12}^{(4)}|x_{21}^{(4)})P(y_{32}^{(4)} |x_{23}^{(4)})\\
P(x_{21}^{(5)} x_{31}^{(5)} y_{12}^{(5)} y_{13}^{(5)})&=P(x_{21}^{(5)})P(x_{31}^{(5)}|x_{21}^{(5)})P(y_{12}^{(5)}|x_{21}^{(5)})P(y_{13}^{(5)}|x_{31}^{(5)})\\
P(x_{13}^{(6)} x_{23}^{(6)} y_{31}^{(6)} y_{32}^{(6)})&=P(x_{13}^{(6)}|x_{23}^{(6)})P(x_{23}^{(6)})P(y_{31}^{(6)}|x_{13}^{(6)})P(y_{32}^{(6)}|x_{23}^{(6)})
\end{align*}
are achievable with a 6P scheme and a partial-decode-and-forward (PDF) strategy.
\end{theorem}
\begin{proof}
In Theorem \ref{prop:AR_6P_PDF} set the channel outputs to
\begin{align}
Y_{1}^{(l)}&=(Y_{12}^{(l)}Y_{13}^{(l)})\notag\\
Y_{2}^{(l)}&=(Y_{21}^{(l)}Y_{23}^{(l)})\notag\\
Y_{3}^{(l)}&=(Y_{31}^{(l)}Y_{32}^{(l)}),\quad l=1,\ldots, 6,
\end{align}
and the channel inputs to
\begin{align}
U_{1}^{(1)}&=X_{12}^{(1)}&X_{1}^{(1)}&=X_{13}^{(1)}\notag\\
U_{3}^{(2)}&=X_{32}^{(2)}&X_{3}^{(2)}&=X_{31}^{(2)}\notag\\
X_{1}^{(3)}&=X_{12}^{(3)}&X_{3}^{(3)}&=X_{32}^{(3)}\notag\\
X_{2}^{(4)}&=(X_{21}^{(4)}X_{23}^{(4)})\notag\\
X_{2}^{(5)}&=X_{21}^{(5)}&X_{3}^{(5)}&=X_{31}^{(5)}\notag\\
X_{2}^{(6)}&=X_{23}^{(6)}&X_{1}^{(6)}&=X_{13}^{(6)}.
\end{align}
Then consider the properties of orthogonality and their effect on the mutual informations while noticing that the sum-rate constraint is no longer binding.
\end{proof}
A more abstract form of Theorem \ref{prop:AR_6P_PDF_ORT} can be given by further simplifying the network to links supporting a certain capacity $b_{ij}^{(l)}$.
\begin{theorem}
\label{prop:AR_6P_PDF_WL}
All rate pairs of the half-duplex two-way relay channel with directed links of capacity $b_{ij}^{(l)}$ that satisfy
\begin{align*}
R_{1\text{-}3}\leq &\tau_{1}\Big(b_{12}^{(1)}+b_{13}^{(1)}\Big)+ \tau_{3}b_{12}^{(3)} + \tau_{6}b_{13}^{(6)}\notag\\
R_{1\text{-}3}\leq &\tau_1 b_{13}^{(1)} + \tau_4 b_{23}^{(4)} + \tau_6 \Big(b_{13}^{(6)}+b_{23}^{(6)}\Big)\notag\\
R_{3\text{-}1}\leq &\tau_{2}\Big(b_{32}^{(1)}+b_{31}^{(2)}\Big)+\tau_{3}b_{32}^{(3)} +\tau_{5}b_{31}^{(5)}\notag\\
R_{3\text{-}1}\leq &\tau_2 b_{31}^{(2)} + \tau_4 b_{21}^{(4)} + \tau_5 \Big(b_{21}^{(5)}+b_{31}^{(5)}\Big)
\end{align*}
with $0<\tau_{l}$ and $\sum_{l=1}^{6}\tau_{l}\leq1$, are achievable with a 6P scheme and a partial-decode-and-forward strategy.
\end{theorem}
\begin{proof}
Follows directly from Theorem \ref{prop:AR_6P_PDF_ORT} after replacing $I(X_{ij}^{(l)};Y_{ji}^{(l)})$ by $b_{ij}^{(l)}$.
\end{proof}
The last two results coincide with the rates of Theorem \ref{prop:UB_6P} applied to the orthogonal case and therefore PDF achieves capacity under these circumstances similar to the one-way relay channel \cite{Gam05}. Note that orthogonality in the last three phases is not essential for communication at channel capacity. 
\section{Linear Resource Allocation Problems}
Assuming fixed input distributions the mutual informations of the outer and inner bounds (Theorem \ref{prop:UB_6P}, \ref{prop:AR_6P_PDF} and \ref{prop:AR_6P_DF}) are constant and only the right time-allocation $\tau_l$ for $l=1,\ldots,L$ has to be found. Fortunately, for various reasonable objectives this problem can be stated as a small-scale linear program which can be solved at low complexity using the simplex algorithm \cite{Bert99}. Possible rate objectives to be maximized are for example
\begin{align}
R_{\text{SRMAX}}&=\max\limits_{\ve{\tau}} \lbrace{R_{1\text{-}3}+R_{3\text{-}1}}\rbrace \notag\\
R_{\text{WSRMAX}}&=\max\limits_{\ve{\tau}}\lbrace{\lambda R_{1\text{-}3}+(1-\lambda)R_{3\text{-}1}}\rbrace\hspace{1cm}\lambda\in[0,1] \notag\\
R_{\text{MAXMIN}}&=\max\limits_{\ve{\tau}}\lbrace{\min (R_{1\text{-}3},R_{3\text{-}1})}\rbrace.
\end{align}
If one is interested in communicating with certain rates $(R_{1\text{-}3}, R_{3\text{-}1})$ at lowest cost, one minimizes
\begin{align}
C_{\text{TCMIN}}=\min\limits_{\ve{\tau}} C_{T}(R_{1\text{-}3},R_{3\text{-}1})
\end{align}
by a linear program after having associated a cost $c_l$ with each phase $l$ of unit duration\footnote{Note that this requires the transmission cost to be a linear function in $\tau$.}. Solving the time-allocation problem can result in some phases to be turned off and therefore gives the optimal scheme for the considered relaying strategy. Note that changing the order of the six phases or splitting them up and permuting the individual parts in an arbitrary way will not result in a better performance as the problem does not allow the source encoders to establish a cooperation and all nodes have agreed on the scheme a priori.
\subsection{Relevance for Wireless Systems}
For \emph{time-division duplex} (TDD) systems the allocation of time-lengths to the individual phases can be used in order to maximize one of the mentioned rate objectives while the channel is used all the time. An example for a cost problem is to minimize the overall duration of the two-way communication for certain rates in order to allow the nodes to perform other communication tasks for the rest of the time. The two-way relaying problem has been stated under the assumption that the nodes are not allowed to transmit and receive at the same time.  Although a rigorous proof would require \emph{block-Markov} arguments as used in \cite{Cov79}, the achievable rate expressions hold also for \emph{frequency-division duplex} (FDD) systems where the different communication parts are carried out in parallel on orthogonal frequencies \cite{Ho05}. For band-limited Gaussian channels \cite[Section 9.3]{Cov06}, $\tau_l$ can be reinterpreted as the bandwidth $\omega_l$ associated with the $l$-th network state. Therefore, one might want to maximize a rate objective by allocating the optimal bandwidth to each of the communication parts. A reasonable cost problem is to minimize the overall used bandwidth for given rates.
\section{Gaussian Two-Way Relay Channel}
Now we consider a scalar \emph{AWGN}-channel model where an active output at node $j$ in phase $l$ has the form 
\begin{align}
Y_j^{(l)}&= \sum_{i=1, i\neq j}^{3}h_{ij}X_i^{(l)}+Z_j^{(l)}
\end{align}
with $X_i^{(l)},Y_j^{(l)},h_{ij}\in\mathbb{C}$ and the additive complex Gaussian noise $Z_{j}^{(l)}$ being independent, zero mean and of unit variance, i.e., $Z_{j}^{(l)}\sim\mathcal{N}_{\mathbb{C}}(0,1)$. The active input distributions are limited by a \emph{per-symbol power constraint}
\begin{align}
\ex{\norm{X_{i}^{(l)}}^{2}} \leq P_i\hspace{1cm}\forall i=1,2,3.
\end{align}
As for such a channel DF performs as good as PDF in the following full decoding (DF) at the relay is used. The input variables have the form
\begin{align}
X_1^{(1)}&=\sqrt{P_1}f_{1}^{(1)}(W_{1\text{-}3})\notag\\
X_3^{(2)}&=\sqrt{P_3}f_{3}^{(2)}(W_{3\text{-}1}) \notag\\
X_1^{(3)}&=\sqrt{P_1}f_{1}^{(3)}(W_{1\text{-}3}) \notag\\
X_3^{(3)}&=\sqrt{P_3}f_{3}^{(3)}(W_{3\text{-}1}) \notag\\
X_2^{(4)}&=\sqrt{P_2}f_{2}^{(4)}(W_{1\text{-}3},W_{3\text{-}1}) \notag\\
X_2^{(5)}&=\sqrt{P_2}f_{2}^{(5)}(W_{3\text{-}1}) \notag\\
X_3^{(5)}&=\underbrace{\sqrt{\beta P_3}f_{2}^{(5)}(W_{3\text{-}1})}_{\sqrt{\beta P_3/P_2} X_2^{(5)}}+\sqrt{(1-\beta)P_3}f_{3}^{(5)}(W_{3\text{-}1}) \notag\\
X_2^{(6)}&=\sqrt{P_2}f_{2}^{(6)}(W_{1\text{-}3}) \notag\\
X_1^{(6)}&=\underbrace{\sqrt{\gamma P_1}f_{2}^{(6)}(W_{1\text{-}3})}_{\sqrt{\gamma P_1/P_2}X_2^{(6)}}+\sqrt{(1-\gamma)P_1}f_{1}^{(6)}(W_{1\text{-}3})
\end{align}
with encoding functions $f_{i}^{(l)}\sim\mathcal{N}_{\mathbb{C}}(0,1)$ while $\beta,\gamma\in[0;1]$ control coherent signaling of nodes 3 and 1 with node 2 in the fifth and sixth phase of the communication protocol. In accordance with theorem \ref{prop:AR_6P_DF} the achievable rates are
\begin{align}
R_{1\text{-}3} &\leq \tau_{1}\loga{1+\norm{h_{12}}^2P_1}+\tau_3\loga{1+\norm{h_{12}}^2P_1}+\tau_6\loga{1+\norm{h_{13}}^2(1-\gamma)P_1} \notag\\
R_{1\text{-}3} &\leq \tau_{1}\loga{1+\norm{h_{13}}^2P_1}+\tau_4\loga{1+\norm{h_{23}}^2P_2}+ \notag\\
&+\tau_{6}\loga{1+\norm{h_{13}}^2P_1+\norm{h_{23}}^2P_2+2\norm{h_{13}h_{23}}\sqrt{\gamma P_1P_2}}\notag\\
R_{3\text{-}1} &\leq \tau_{2}\loga{1+\norm{h_{32}}^2P_3}+\tau_3\loga{1+\norm{h_{32}}^2P_3}+\tau_{5}\loga{1+\norm{h_{31}}^2(1-\beta)P_3} \notag\\
R_{3\text{-}1} &\leq \tau_{2}\loga{1+\norm{h_{31}}^{2}P_3}+\tau_4\loga{1+\norm{h_{21}}^2P_2}+ \notag\\
&+\tau_{5}\loga{1+\norm{h_{31}}^2P_3+\norm{h_{21}}^2P_2+2\norm{h_{31}h_{21}}\sqrt{\beta P_2 P_3}} \notag\\
R_{1\text{-}3}+R_{3\text{-}1} &\leq \tau_{1}\loga{1+\norm{h_{12}}^2P_1}+\tau_{2}\loga{1+\norm{h_{32}}^2P_3}+\tau_3\loga{1+\norm{h_{12}}^2P_1+\norm{h_{32}}^2P_3}+\notag\\
&+\tau_{5}\loga{1+\norm{h_{31}}^2(1-\beta)P_3}+\tau_6\loga{1+\norm{h_{13}}^2(1-\gamma)P_1}.
\end{align}
\section{Simulations}
\begin{figure}[!h]
\centering
\includegraphics[width=8cm]{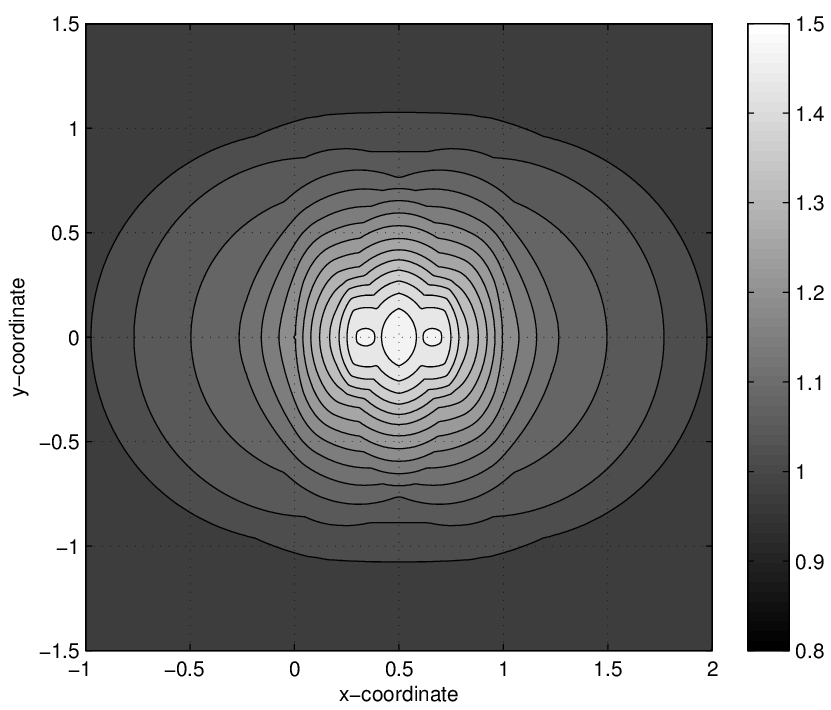}
\caption{$R_{\text{MAXMIN}}$ 6P-DF/TWC}
\label{fig:MM_DF_VS_TWC}
\end{figure}

\begin{figure}[!h]
\centering
\includegraphics[width=8cm]{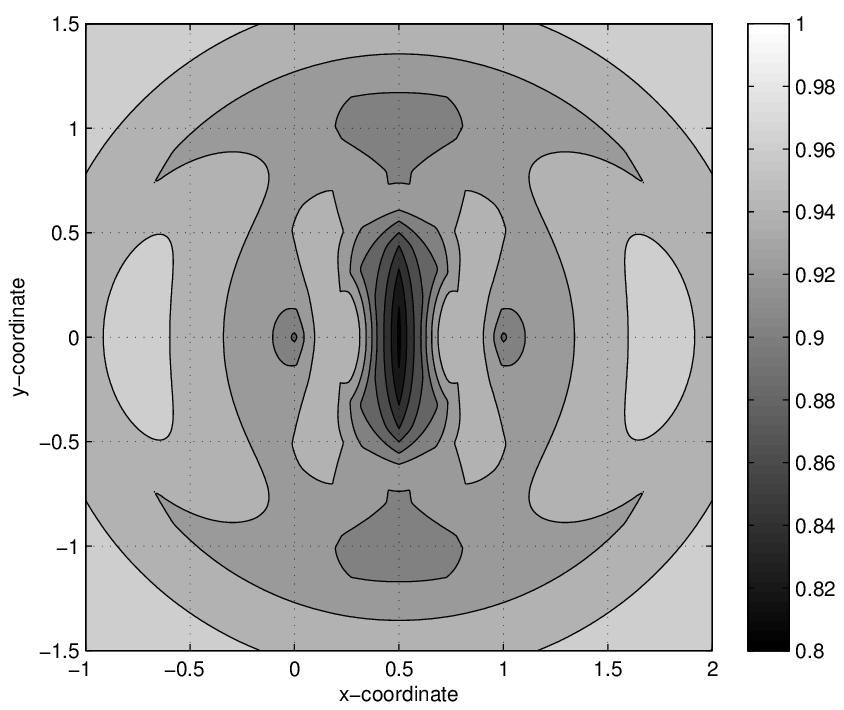}
\caption{$R_{\text{MAXMIN}}$ 6P-DF/UB}
\label{fig:MM_DF_VS_UB}
\end{figure}

\begin{figure}[!h]
\centering
\includegraphics[width=8cm]{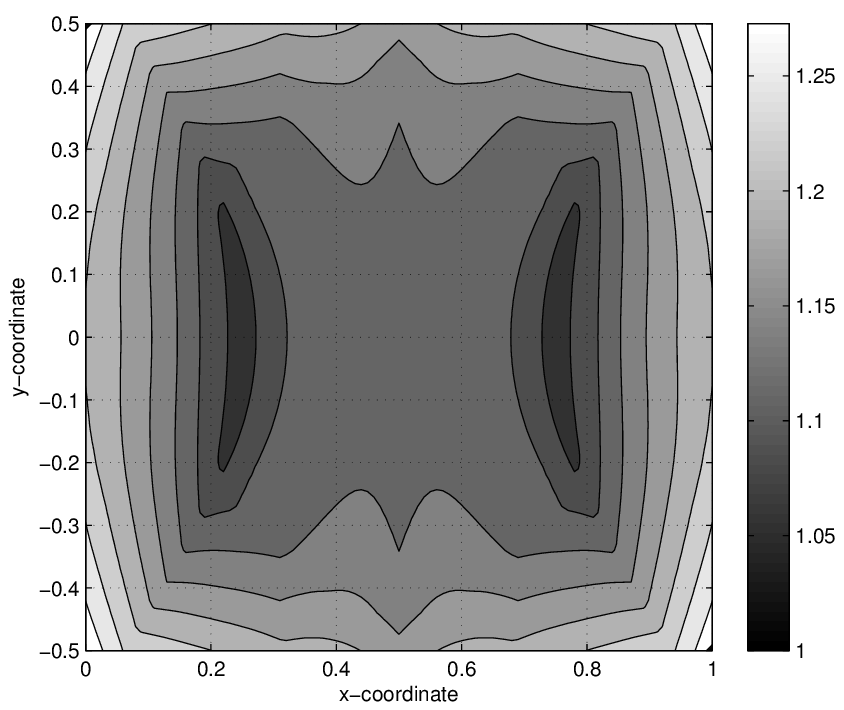}
\caption{$R_{\text{MAXMIN}}$ 6P-DF/2P-DF}
\label{fig:MM_2P_VS_6P}
\end{figure}

For simulations a plane-network model \cite{Kra05} is used where the positions $\mbox{\boldmath$\xi$}_{i}$ of node 1 and 3 are fixed to $\mbox{\boldmath$\xi$}_{1}=\begin{bmatrix}0 &0\end{bmatrix}^T$, $\mbox{\boldmath$\xi$}_{3}=\begin{bmatrix}1 &0\end{bmatrix}^T$ and node 2 can be placed at position $\mbox{\boldmath$\xi$}_{2}=\begin{bmatrix}x &y\end{bmatrix}^T$ with $x,y\in(-\infty,\infty)$. The channel coefficient 
\begin{align}
h_{ij}=1/d_{ij}^{\frac{\alpha}{2}}
\end{align}
is determined by the path-loss exponent $\alpha$ and the distance $d_{ij}=\Vert \mbox{\boldmath$\xi$}_{i} -\mbox{\boldmath$\xi$}_{j} \Vert_{2}$ between node $i$ and node $j$. For a scenario with $\alpha=3$ and a power constraint $P_i=10$ like used in \cite{Rank06}, we move the relay to different positions, sample over the parameters $\beta,\gamma$ and solve time allocation with the objective of maximizing the symmetric  two-way rate $R_{\text{MAXMIN}}$. After sampling we pick $\beta^{\star},\gamma^{\star}$ in conjunction with the time allocation solution $\ve{\tau}^{\star}$ that yield the highest symmetric rate. Note that if, due to system constraints, coherent signaling of two nodes is not possible one sets $\beta=\gamma=0$ and solves the problem at each position by one linear program. Fig. \ref{fig:MM_DF_VS_TWC} shows the achievable rate gain compared to two-way communication without relay. It can be observed that in the area between the nodes an increase in symmetric rate of $25$ to $50$ percent is possible. Fig. \ref{fig:MM_DF_VS_UB} compares the achievable symmetric rates with their upper bound. This shows that with the relay being located in the middle between the two dialog nodes still an improvement of $20$ percent might be possible with other relaying strategies. For the other regions DF performs within $10$ percent distance to the upper bound. A comparison against the two-phase approach (see Fig. \ref{fig:2WRC2STEPS}) is given in Fig. \ref{fig:MM_2P_VS_6P}. The suggested strategy outperforms this transmission protocol by about $5$ to $15$ percent when the relay is located between the dialog nodes. As the two-phase strategy neglects the direct path, the rate gain grows continuously towards infinity if the relay moves away from the dialog nodes.
\section{Conclusion}
The problem of exchanging messages between two nodes in a fully-connected half-duplex three-node network has been defined and an outer bound on the achievable rates has been derived. This bound alludes to a scheme that takes into account all possible network states that are relevant. For such a scheme coding procedures with partial and full decoding at the relay have been proposed. The coding proof is based on sequential decoding of multidimensional message indices at the dialog nodes, giving insides into a message passing structure that might guide finite block-length code constructions for the considered communication problem. For fixed input distributions the optimal time-allocation to the individual network states can be found at low complexity. Possible applications of this property for wireless systems have been outlined. Simulations for the established communication protocol show a significant increase in symmetric rate compared to two-way communication without relay and outperformance of strategies that ignore the direct path. 

\section*{Acknowledgment}
The presented results are part of the author's \emph{Diplomarbeit} (equivalent to a Master Thesis) \cite{Stein10}. The thesis has been conducted from November 2009 till October 2010 at the Associate Institute for Signal Processing, Technische Universit\"at M\"unchen.

\ifCLASSOPTIONcaptionsoff
  \newpage
\fi

\bibliographystyle{IEEEtran}

\end{document}